\documentclass[floatfix,twocolumn,showpacs,preprintnumbers,amsmath,amssymb,superscriptaddress]{revtex4-1}

\usepackage{graphicx}
\usepackage{dcolumn}
\usepackage{bm}
\usepackage{amsthm,amsmath}

\usepackage{color}

\usepackage{tikz}
\usepackage[T1]{fontenc}
\usepackage{lmodern}
\tikzstyle{format}=[draw, thin, rounded corners=5pt, line width =0.5, fill=concepts!10!white]
\usetikzlibrary{patterns,arrows}
\usetikzlibrary{decorations.pathreplacing,decorations.markings,decorations.pathmorphing,snakes}
\def\myscale{1.0}
\def\myscaletwo{0.5}
\def\pta{5.0}
\def\ptb{9.0}

\begin{document}

\preprint{AIP/123-QED}

\title[GOA]{Wave-function inspired density functional applied to the H$_2$/H$_2^+$ challenge}

\author{Igor Ying Zhang}
\affiliation{Fritz-Haber-Institut der Max-Planck-Gesellschaft, Faradayweg 4-6, 14195 Berlin, Germany}%
\email{zhang@fhi-berlin.mpg.de}

\author{Patrick Rinke}%
\affiliation{Fritz-Haber-Institut der Max-Planck-Gesellschaft, Faradayweg 4-6, 14195 Berlin, Germany}%
\affiliation{Department of Applied Physics, Aalto University, P.O. Box 11100, Aalto FI-00076, Finland}%


\author{Matthias Scheffler}
\affiliation{Fritz-Haber-Institut der Max-Planck-Gesellschaft, Faradayweg 4-6, 14195 Berlin, Germany}%
\affiliation{Department of Chemistry and Biochemistry and Materials Department, University of California-Santa Barbara, Santa Barbara, CA 93106-5050, USA}%

\date{\today}

\begin{abstract}

	We start from the Bethe-Goldstone equation (BGE) to derive a simple orbital-dependent correlation functional -- BGE2 --
	which terminates the BGE expansion at the second-order, but retains the self-consistent coupling of electron-pair
	correlations.
	We demonstrate that BGE2 is size consistent and one-electron ``self-correlation'' free. The electron-pair correlation 
	coupling ensures the correct H$_2$ dissociation limit and gives a finite correlation energy for any system even if it has a
	no energy gap. BGE2 provides a good description of	both H$_2$ and H$_2^+$ dissociation, which is regarded as a great challenge in density functional theory (DFT). 
	We illustrate the behavior of BGE2 analytically by considering H$_2$ in a minimal basis. Our analysis shows that BGE2
	captures essential features of the adiabatic connection path that current state-of-the-art DFT approximations do not.
\end{abstract}

\pacs{Valid PACS appear here}
\maketitle

\section{Introduction}
Density-functional theory (DFT) is now widely applied in physics, chemistry, materials science and biology. 
This success comes from the availability of suitable approximations for the exchange-correlation ($xc$) 
functional -- the only quantity that is unknown in Kohn-Sham (KS) DFT. However, despite this unmatched success and the ubiquitous 
application of common functionals, all currently available functionals suffer from certain notorious limitations.
For example, all existing functionals fail to correctly describe the dissociations of both
H$_2^+$ \emph{and} H$_2$, two very simple molecules\cite{yang:2011A,gorling:2011A,furche:2013A,scuseria:2010A,scuseria:2010B}. 
Solving the H$_2^+$/H$_2$ dissociation problem will not only lead to a conceptual understanding 
of why current functionals fail, but also offer potential pathways to develop better functionals, and has
thus attracted increasing attention\cite{ernzerhof:1996A,perdew:2000A,burke:2005A,becke:2005A,yang:2007A,perdew:2008A,gorling:2008A,perdew:2010A,
scuseria:2010A,scuseria:2010B,becke:2013A,gorling:2011A,rinke:2013A,rinke:2013B,scuseria:2010A,toulouse:2012A,scuseria:2013A,furche:2013A,olsen:2014A}. 

A viable approach in functional development is to learn from wave-function theory in constructing nonlocal correlation functionals 
that  involve unoccupied KS orbitals and thus stand on the fifth (and currently highest) rung of Perdew's Jacob's ladder\cite{perdew:2000A}.
A prominent example is second-order G\"orling-Levy perturbation theory (GL2)\cite{gorling:1994A} that is closely related to second-order
M\o{}llet-Plesset perturbation theory (MP2)\cite{plesset:1934A,fetter:1996A,szabo:1996A} in wave-function theory\cite{singles}. In fact, 
even MP2 can be viewed as an implicit density functional by means of the adiabatic connection approach\cite{perdew:1977A,gorling:1993A,gorling:1994A}. 
An important feature of 2nd-order perturbation theories (PT2) such as GL2 and MP2 is that they are one-electron ``self-correlation'' 
free\cite{yang:2011A,scuseria:2010B}, i.e.\ the correlation is zero for one-electron systems, and that they are size consistent (i.e. if 
the system is fragmented into two parts, the total energy becomes the sum of the two fragments) \cite{bartlett:1981A,hirata:2011A}. 
Furthermore, PT2 is fully non-local and captures the correct R$^{-6}$ decay behaviour at long distances, which is essential to provide an 
accurate description of weak interactions. Therefore, PT2 correlation is an ideal building block for 
fifth-level density functionals, following Perdew's nomenclature. This feature has been exploited in a range of promising double-hybrid functionals\cite{grimme:2006A,martin:2008A,
igor:2009A,chai:2009A,grimme:2011A,martin:2011A,toulouse:2011A,igor:2011B,igor:2013B}, which linearly mix generalized 
gradient approximations (GGAs), e.g. BLYP\cite{becke:1988A,yang:1988A} or PBE\cite{perdew:1996A}, with both exact exchange 
and PT2 correlation. The admixture of semi-local exchange and correlation can be viewed as an efficient yet semi-empirical way to take higher-order perturbative contributions into account that would go beyond PT2\cite{igor:2011A,igor:2014A}. These 
double-hybrid functionals provide a satisfactory accuracy for various chemical interactions, but they fail for systems with 
small KS energy gaps, e.g. heavily-stretched H$_2$ and metallic systems. In such systems, two or more determinants become 
degenerate in energy and mean-field theories that rely on a single reference determinant such as HF or KS-DFT break down. 
Also,  perturbation theory diverges at any order, making it essential to find an appropriate resummation. 
Seidl, Perdew and Kurth suggested an empirical adiabatic-connection (AC) model\cite{perdew:1977A,gorling:1993A,gorling:1994A}, namely interaction-strength 
interpolation (ISI)\cite{perdew:2000A}, to implicitly resum the perturbation expansion by using only exact exchange, PT2 
correlation and an explicit density functional derived from the "point charge plus continuum" model in the strong-interaction 
limit where the coupling constant parameter goes to infinity\cite{perdew:2000B}. Frequently, the analytic dependence on the 
coupling constant is approximated by a Pad\'e formula, whose  parameters can be determined either empirically on theoretical
grounds\cite{ernzerhof:1996A,yang:2007A}. Recently, an explicit density functional in the strong-interaction limit was suggested. 
It can be constructed from so-called co-motion functions and captures fully non-local effects in the strong-interaction limit\cite{paola:2012A}.

An example of a successful resummation is the particle-hole random-phase approximation (RPA), in which an infinite number of ``ring diagrams''
is summed \cite{perdew:1975A,furche:2001A,burke:2005A,jiang:2007A,kresse:2008A,rinke:2009A,igor:2015A}. This has 
been widely recognized as key to make RPA applicable to small-gap or metallic systems. The RPA method provides the correct H$_2$ 
dissociation limit \cite{gorling:2011A,rinke:2013A}. Unfortunately, it suffers from a heavy ``self-correlation'' error for one-electron 
systems\cite{yang:2011A,gorling:2011A,furche:2013A,scuseria:2010A,scuseria:2010B}, and thus yields an even worse H$_2^+$ dissociation
behaviour than conventional density functionals. In addition,  RPA exhibits an incorrect repulsive ``bump'' at intermediate H$_2$ bond
distances. These deficiencies have in the past been attributed to a lack of self-consistency in RPA \cite{scuseria:2010B,gorling:2011A}, 
which was disproved by actual self-consistent 
calculations\cite{gross:2012A,rinke:2013A}. Another widely accepted hypothesis attributes these deficiencies to the lack of higher order
diagrams and spurred considerable beyond-RPA developments in the past few years\cite{rinke:2011A,rinke:2013B,scuseria:2008A,bartlett:2011A,toulouse:2011B,
scuseria:2013A,yang:2013A,yang:2014A,furche:2013A,olsen:2014A}. Other interesting developments in this realm include reduced density matrix theory \cite{pernal:2013A} or self-consistent Green's function frameworks\cite{rinke:2013A,phillips:2014A}.
A successful beyond-RPA method is renormalized second-order perturbation theory (rPT2), which adds an infinite summation of the second-order
exchange diagram of PT2 (termed second order screened exchange (SOSEX))\cite{Freeman:1977,kresse:2010A,scuseria:2010A,Paier/etal:2012} and
renormalized single-excitation (rSE)  \cite{rinke:2011A,RPAreview} diagrams on top 
of RPA\cite{rinke:2013B}. rPT2 does not diverge for small-gap systems and is free of one-electron ``self-correlation''. It thus
provides the correct description of one-electron systems including H$_2^+$ and individual H atoms, but fails for H$_2$ dissociation if breaking spin symmetry is not allowed [Refs. \onlinecite{rinke:2013B,furche:2013A,scuseria:2008A} 
and also below]. Further improvements have been stipulated in the context of the couple-cluster (CC)  theory\cite{scuseria:2008A,bartlett:2011A,toulouse:2011B,scuseria:2013A,yang:2013A,yang:2014A} or the Bethe-Salpeter equation \cite{fetter:1996A,furche:2013A,olsen:2014A}. These methods, although proposed from different perspectives, 
can all be interpreted as attempts to explicitly introduce more ``selective summations to infinite order'' in the density 
functional perturbation framework. Even though these methods improve over standard RPA schemes with varying degrees of success for the H$_2^+$/H$_2$ dissociation problem, no improvement to date removes the ``bump'' while simultaneously yielding the correct dissociation limit for H$_2^+$ and H$_2$, indicating the difficulty of understanding this problem in any perturbative framework.

In this paper, we lay the ground for an efficient orbital-dependent correlation functional based on the Bethe-Goldstone equation 
(BGE)\cite{bethe:1957A}, which is derived from the correlation of two particles\cite{fetter:1996A}. As the BGE is the simplest
approximation which provides the exact solution for one- and two-electron systems, it is a good starting point to understand the 
aforementioned H$_2^+$/H$_2$ dissociation challenge. In Sec.\ \ref{Sec:BGE}, we formulate the BGE in the context of DFT through the 
adiabatic connection approach. In contrast to the normal resummation strategy in density functional perturbation theory, we propose 
a new correlation functional by terminating the BGE expansion at the second order (BGE2). As shown in Sec.\ \ref{Sec:BGE2}, this BGE2 approximation gives a good description of both H$_2^+$ and H$_2$ dissociations, without requiring any higher order connected
Goldstone diagrams that are commonly believed to be necessary. We further analyse BGE2 analytically in the minimal basis
H$_2$ model. We show that BGE2 is size-extensive and free of one-electron ``self-correlation''.

\section{\label{Sec:BGE} Bethe-Goldstone equation in density functional theory}

In the adiabatic-connection (AC) approach of density functional theory (DFT)\cite{perdew:1977A,gorling:1993A,gorling:1994A}, the 
non-interacting KS system is connected to the physical system by an adiabatic path. The density  $n$ along the path  is fixed to the exact 
ground-state density. The Hamiltonian for a family of partially interacting $N$-electron systems in this path is controlled by a 
coupling-constant parameter $\lambda$, (atomic units are used hereafter unless stated otherwise):
\begin{equation}
	\label{Eq:Hlambda}
	\hat{H}_{\lambda}
	=\hat{H}_{s}+\lambda(\hat{V}_{ee}-\hat{v}_{\lambda}/\lambda)
	=\hat{H}_{s}+\lambda \Delta_{\lambda}
\end{equation}
Here, $\hat{H}_{s}$ is the Hamiltonian of the non-interacting KS system 
\begin{equation}
	\label{Eq:Hs}
	\hat{H}_{s}=\sum_{i}^{N}\left[-\frac{1}{2}\nabla_i^2+v_{s}(\boldsymbol{r}_i)\right]
\end{equation}
where $v_{s}(\boldsymbol{r})$ is a multiplicative one-electron potential
\begin{equation}
	v_{s}(\boldsymbol{r})=v_{\textrm{ext}}(\boldsymbol{r})+v_{\rm H}(\boldsymbol{r})+v_{x}(\boldsymbol{r})+v_{c}(\boldsymbol{r})
\end{equation}
comprising the external potential ($v_{\textrm{ext}}$) arising from the Coulomb interaction between the electrons and the nuclei, the Hartree potential ($v_{\rm H}$), and the exchange ($v_{x}$) and correlation  ($v_{c}$) potential. The operator $\hat{v}_{\lambda}$ is also multiplicative and constrained to satisfy $\hat{v}_0=0$ and $\hat{v}_{1}=\hat{v}_{H}+\hat{v}_{xc}$. Thus, $\hat{H}_{\lambda=0}=\hat{H}_{s}$, while $\hat{H}_1$ is the Hamiltonian of the fully interacting system. From  the perspective of many-body perturbation theory, $\Delta_{\lambda}=\hat{V}_{ee}-\hat{v}_{\lambda}/\lambda$ is a perturbation of the non-interaction KS Hamiltonian, which does not change the ground-state density $n$. By using coordinate scaling\cite{levy:1991A,gorling:1993A,gorling:1994A}, it was shown that
\begin{equation}
	\label{Eq:CS}
	\begin{split}
	    \hat{v}_{\lambda}/\lambda = \sum_{i=1}^{N}&\left[v_{H}(\boldsymbol{r_i})+v_{x}(\boldsymbol{r_i})+\lambda\frac{\delta E_{c}[n_{\alpha}]}
	    {\delta n(\boldsymbol{r}_i)}\right]\textrm{, }\alpha=\lambda^{-1}\\
		&v_{c}(\boldsymbol{r}_i,\alpha) \overset{!}{=}\frac{\delta E_{c}[n_{\alpha}]}{\delta n(\boldsymbol{r}_i)}
	\end{split}
\end{equation}
where $n_{\alpha}(x,y,z)=\alpha^3n(\alpha x,\alpha y,\alpha z)$, and $v_{c}(\boldsymbol{r}_i,\alpha)$ is the correlation potential of the 
scaled correlation energy with respect to the normal density $n$.

In the AC framework\cite{perdew:1977A,gorling:1993A,gorling:1994A}, the $xc$ functional can be interpreted 
as the coupling-constant integration,
\begin{equation}
	\label{Eq:AC}
	E_{xc}[n]=\int_0^1 d\lambda\frac{\partial E_{xc}^{\lambda}[n]}{\partial \lambda}=\int_0^1 d\lambda V_{xc}^{\lambda}
\end{equation}
where $E_{xc}^{\lambda}[n]$ is the $xc$ functional for a given coupling-constant $\lambda$\cite{perdew:2003A}
\begin{equation}
	\begin{split}
		E_{xc}^{\lambda}[n]
		=&\left<\Psi_{n}^{\lambda}\left|\hat{T}+\lambda\hat{V}_{ee}\right|\Psi_{n}^{\lambda}\right>-
		\left<\Phi_{n}\left|\hat{T}\right|\Phi_{n}\right>-\lambda E_{H}[n].
	\end{split}
\end{equation}
$V_{xc}^{\lambda}$ is the corresponding $xc$ potential for a given coupling constant $\lambda$. We can further define the exchange $E_{x}^{\lambda}[n]$ and correlation $E_{c}^{\lambda}[n]$ components 
separately
\begin{equation}
	\label{Eq:xc-lambda}
	\begin{split}
		E_{x}^{\lambda}[n]=&\lambda\left(\left<\Phi_{n}\left|\hat{V}_{ee}\right|\Phi_{n}\right>-E_{H}[n]\right)=\lambda E_{x}[n]\\
		E_{c}^{\lambda}[n]=&\left<\Psi_{n}^{\lambda}\left|\hat{H}_{\lambda}\right|\Psi_{n}^{\lambda}\right>
		-\left<\Phi_{n}\left|\hat{H}_{\lambda}\right|\Phi_{n}\right>.
	\end{split}
\end{equation}
Here $\Psi_{n}^{\lambda}$ is the ground-state wave-function on the AC path with the coupling constant 
$\lambda$, which gives the same ground-state density $n$ as the physical system ($\lambda=1$). $\Psi_{n}^{0}=\Phi_{n}$ is thus
the ground-state wave-function of the non-interacting KS system. $E_{H}[n]$ is the Hartree energy
\begin{equation}
	E_{H}[n]=\frac{1}{2}\int d\boldsymbol{r}_1d\boldsymbol{r}_2\frac{n(\boldsymbol{r}_1)n(\boldsymbol{r}_2)}
	{\left|\boldsymbol{r}_1-\boldsymbol{r}_2\right|}
\end{equation}
which is an explicit functional of the density. Immediately, we have $V_{x}^{\lambda}=E_{x}[n]$, as the exchange density functional 
$E_{x}^{\lambda}[n]$ defined in this manner is linear in the coupling constant $\lambda$. And the corresponding Hartree potential $v_{H}$ is written as
\begin{equation}
\left<\phi_{a}|v_{H}|\phi_{a}\right>=\sum_{i}^{occ}\left<\phi_a\phi_i|\phi_a\phi_i\right>
\end{equation}
with the definition of the two-electron four-center integral as
\begin{equation}
	\begin{split}
\left<\phi_i\psi_j|\phi_k\phi_l\right>&=\int d\boldsymbol{r}_1d\boldsymbol{r}_2
\frac{\phi_i^*(\boldsymbol{r}_1)\psi_j^*(\boldsymbol{r}_2)\phi_k(\boldsymbol{r}_1)\phi_l(\boldsymbol{r}_2)}{\left|\boldsymbol{r}_1-\boldsymbol{r}_2\right|}.
\end{split}
\end{equation}

In contrast, it is in general not possible to obtain the exact $E_{xc}^{\lambda}[n]$ for any $\lambda\neq 0$, since the electron-electron 
repulsion operator $\hat{V}_{ee}$ appears explicitly in the Hamiltonian, and the ground-state wave-function $\Psi_{n}^{\lambda}$
cannot be obtained exactly. This is also true for two-electron systems, although the ground-state wave-function is now just a simple electron-pair function 
\begin{equation}
	\label{Eq:pair0}
	\Psi_{n}^{\lambda}=\Psi_{ab}.
\end{equation}
As one of the motivations in this paper is to construct a functional which can provide an accurate description for both H$_2$
and H$_2^+$ dissociations, we start from the Bethe-Goldstone equation (BGE) of $\hat{H}_{\lambda}$\cite{fetter:1996A}, which
is derived from the correlation of two particles, and is thus the exact solution for one- and two-electron systems. The BGE 
explicitly solves the Schr\"odinger equation for each electron pair $ab$ interacting through a perturbation $\hat{H}_1(\lambda)$
\begin{equation}
	\label{Eq:pair}
	\begin{split}
		\left[ \hat{H}_{s}+\lambda\hat{H}_1(\lambda) \right]\Psi_{ab} &= E_{ab}\Psi_{ab}\\
		\left[E_{ab}-\hat{H}_{s}\right]\Psi_{ab} &= \lambda\hat{H}_1(\lambda)\Psi_{ab}.
	\end{split}
\end{equation}
Here, we consider the electron-electron interaction $\hat{V}_{ee}$ of electron pair $ab$ explicitly, while leaving the interaction 
with the other $N$-2 electrons on the mean field level $\hat{v}_{ab}^{\textrm{MF}}$. For two electrons we trivially have $\hat{v}_{ab}^{\textrm{MF}}=0$. 
However, for more than two electrons we would have to make this approximation explicitly. The resulting perturbation operator is
\begin{equation}
	\hat{H}_{1}(\lambda)=\hat{V}_{ee}-\hat{v}_{\lambda}/\lambda + \hat{v}_{ab}^{\textrm{MF}}
\end{equation}
with the definition of $\hat{v}_{ab}^{\textrm{MF}}$ as
\begin{equation}
	\label{Eq:MF}
	\begin{split}
		\left<\Phi_{ab}|\hat{v}_{ab}^{\textrm{MF}}|\Phi_{ab}\right>&=\frac{1}{2}\sum_{i=a,b}\sum_{j\ne a,b}^{occ}\left<\phi_i\phi_j||\phi_i\phi_j\right>
	\end{split}
\end{equation}
where $\{\phi_{i}\}$ are the KS orbitals and $\left<\phi_i\phi_j||\phi_i\phi_j\right>=\left<\phi_i\phi_j|\phi_i\phi_j\right>-\left<\phi_i\phi_j|\phi_j\phi_i\right>$. 
The KS orbitals can be used to generate an antisymmetric non-interacting KS electron-pair function 
\begin{equation}
	\Phi_{ab}(1,2)=\frac{1}{\sqrt{2}}\left|
	\begin{array}{cc}
		\phi_{a}(1) & \phi_{b}(1) \\
		\phi_{a}(2) & \phi_{b}(2) \\
	\end{array}
	\right|.
\end{equation}
Here the numbers $1$ and $2$ are a short-hand notation for the tuple of space and spin variables of the first and second electron, respectively.
The corresponding non-interacting Green's function $G_{0}(1,2;1',2';E_{ab})$ for this electron pair $ab$ is 
\begin{widetext}
\begin{equation}
	\label{Eq:GF0_1}
	    G_{0}(1,2;1',2';E_{ab})=\left(\frac{\Phi_{ab}(1,2)}{E_{ab}-\epsilon_a-\epsilon_b}
		+\sum_{i=a,b}\sum_{r}^{unocc}\frac{\Phi_{ir}(1,2)}{E_{ab}-\epsilon_i-\epsilon_r}
	    +\sum_{r<s}^{unocc}\frac{\Phi_{rs}(1,2)}{E_{ab}-\epsilon_r-\epsilon_s}\right)\Psi_{ab}^{*}(1',2')
\end{equation}
\end{widetext}
where $\left\{ \epsilon_i \right\}$ are the KS eigenvalues
\begin{equation}
	\epsilon_i=\left<\phi_i\left|\hat{H}_{s}\right|\phi_i\right>
	=\left<\phi_i\left|\hat{T}\right|\phi_i\right>+\left<\phi_i\left|v_{s}\right|\phi_i\right>.
\end{equation}
As $\hat{H}_{s}$ (Eq.\ \ref{Eq:Hs}) is a one-electron operator, it is also possible to reorganize the eigenvalues in terms of each 
electron pair $ab$
\begin{equation}
	\label{Eq:non-interacting-pair}
	\epsilon_{ab}=\epsilon_a+\epsilon_b = \left<\Phi_{ab}\left|\hat{H}_{s}\right|\Phi_{ab}\right>
\end{equation}
which could be considered as the zero-order approximation of the electron pair energy $E_{ab}$.

Now we introduce the first approximation to the BGE. We neglect the single excitation contribution, i.e., the first sum in Eq.\ \ref{Eq:GF0_1}. 
\begin{equation}
	\label{Eq:GF0}
      \begin{split}
	    G_{0}&(1,2;1',2';E_{ab}) \approx  \\
	    & \left(\frac{\Phi_{ab}(1,2)}{E_{ab}-\epsilon_a-\epsilon_b}
	    +\sum_{r<s}^{unocc}\frac{\Phi_{rs}(1,2)}{E_{ab}-\epsilon_r-\epsilon_s}\right)\Psi_{ab}^{*}(1',2')
      \end{split}
\end{equation}
As will be discussed in Sec.\ \ref{SSec:H2min}, this approximation, together with the other two approximations we will make later, is essential for achieving an efficient correlation functional which, however, keeps the exact solution for the H$_2$ dissociation limit in the minimal basis.

With this approximation, the electron-pair function $\Psi_{ab}$ can be written as \\

\begin{widetext}
\begin{equation}
	\label{Eq:BGE-GF}
	\begin{split}
		\Psi_{ab}(1,2)=&\int d1'd2'G_{0}(1,2;1',2';E_{ab})\lambda \hat{H}_{1}(\lambda)(1',2')\Phi_{ab}(1',2')\\
		=&\Phi_{ab}(1,2)\frac{\left<\Phi_{ab}\left|\lambda \hat{H}_{1}(\lambda)\right|\Psi_{ab}\right>}{E_{ab}-\epsilon_a-\epsilon_b}
		+\sum_{rs}^{unocc}\frac{\Phi_{rs}(1,2)}{E_{ab}-\epsilon_r-\epsilon_s}\left<\Phi_{rs}\left|\lambda \hat{H}_{1}(\lambda)\right|\Psi_{ab}\right>.
	\end{split}
\end{equation}
It is convenient to introduce intermediate normalization  $\left<\Phi_{ab}|\Psi_{ab}\right>=1$.  Together with 
the expression of the expectation value of the perturbation energy
\begin{equation}
	\label{Eq:IN}
	E_{ab}-\epsilon_{a}-\epsilon_{b}=\left<\Phi_{ab}\left|\lambda\hat{H}_{1}(\lambda)\right|\Psi_{ab}\right>
\end{equation}
the BGE electron pair function $\Psi_{ab}(1,2)$ becomes
\begin{equation}
	\label{Eq:BGE-1}
	\begin{split}
		\Psi_{ab}(1,2)
		=&\Phi_{ab}(1,2)
		+\sum_{r<s}^{unocc}\frac{\Phi_{rs}(1,2)}{E_{ab}-\epsilon_r-\epsilon_s}\left<\Phi_{rs}\left|\lambda \hat{H}_{1}(\lambda)\right|\Psi_{ab}\right>.
	\end{split}
\end{equation}
Since the BGE electron pair function $\Psi_{ab}$ appears on both sides of this equation, both $\Psi_{ab}$ and $E_{ab}$ 
(eq.\ \ref{Eq:BGE-2}) contain an infinite sequence of Goldstone diagrams\cite{fetter:1996A,szabo:1996A}, as one can 
easily see by inserting Eq.\ \ref{Eq:BGE-1} into Eq.\ \ref{Eq:IN}
\begin{equation}
	\label{Eq:BGE-2}
	\begin{split}
		E_{ab}-\epsilon_{a}-\epsilon_{b}=&\left<\Phi_{ab}\left|\lambda\hat{H}_{1}(\lambda)\right|\Phi_{ab}\right>
		+\sum_{r<s}^{unocc}\frac{\left<\Phi_{ab}\left|\lambda\hat{H}_{1}(\lambda)\right|\Phi_{rs}\right>
		\left<\Phi_{rs}\left|\lambda\hat{H}_1(\lambda)\right|\Psi_{ab}\right>}
		{E_{ab}-\epsilon_{r}-\epsilon_{s}}.
	\end{split}
\end{equation}
\end{widetext}

We now consider the different terms step by step. First, we expand the $e_{ab}^{1st}(\lambda)$ term on the right-hand side
\begin{equation}
	\label{Eq:BGE-1st}
	\begin{split}
		e_{ab}^{1st}(\lambda)&=\left<\Phi_{ab}\left|\lambda\hat{H}_{1}(\lambda)\right|\Phi_{ab}\right>\\
		&=\lambda\left<\Phi_{ab}\left|\hat{V}_{ee}+\hat{v}^{\textrm{MF}}_{ab}-\hat{v}_{\lambda}/\lambda\right|\Phi_{ab}\right>.
	\end{split}
\end{equation}
It is the first-order correction to the non-interaction electron pair energy $\epsilon_{ab}$ defined in Eq.\ \ref{Eq:non-interacting-pair}. 
Utilizing the definitions of $\hat{v}_{\lambda}$ (Eq.\ \ref{Eq:CS}) and $\hat{v}_{ab}^{\textrm{MF}}$ (Eq.\ \ref{Eq:MF}) for the 
fully-interacting system ($\lambda=1$), we have
\begin{equation}
	\label{Eq:BGE-1st-lambda=1}
	\begin{split}
		e_{ab}^{1st}(1)
		&=\left<\Phi_{ab}\left|\hat{V}_{ee}+\hat{v}^{\textrm{MF}}_{ab}-\hat{v}_{H}-\hat{v}_{xc}\right|\Phi_{ab}\right>\\
		&=\sum_{i=a,b}\left<\phi_{i}\left|\frac{1}{2}(\hat{v}_{x}^{\textrm{HF}}-\hat{v}_{H})-\hat{v}_{xc}\right|\phi_{i}\right>\\
		&=e_{a}^{1st}(1)+e_{b}^{1st}(1)\\
	\end{split}
\end{equation}
where $\epsilon_{a}^{1st}(1)$ is the corresponding first-order correction of the non-interaction electron energy $\epsilon_{a}$ for the fully-interacting system.
And $\hat{v}_{x}^{\textrm{HF}}$ is the Hartree-Fock like exact exchange operator defined as
\begin{equation}
\left<\phi_a|\hat{v}_{x}^{\textrm{HF}}|\phi_a\right>=-\sum_{i}^{occ}
\left<\phi_a\phi_i|\phi_i\phi_a\right>.
\end{equation}

Together with the the non-interaction electron pair energy 
$\epsilon_{ab}$, this leads to the electron-pair total energy at the first-order many-body perturbation level, which contains only the exact exchange.
Next we define the BGE electron-pair correlation energy $e_{ab}^{\textrm{BGE}}(\lambda)$ as
\begin{equation}
	\label{Eq:BGE-eab}
	\begin{split}
		&e_{ab}^{\textrm{BGE}}(\lambda)=E_{ab}-\epsilon_a-\epsilon_b-e_{ab}^{1st}(\lambda)\\
	\end{split}
\end{equation}
For two electrons, $e_{ab}^{\textrm{BGE}}(\lambda)$ is the total correlation energy $E_{c}^{\textrm{BGE}}[n](\lambda)$. For more than two electrons we 
have to sum up the the correlation energies of all electron pairs:
\begin{equation}
	\label{Eq:BGE-Eab}
	\begin{split}
		&E_{c}^{\textrm{BGE}}[n](\lambda)=\sum_{a<b}^{occ}e_{ab}^{\textrm{BGE}}(\lambda)
	\end{split}
\end{equation}
Finally, the BGE total energy for the fully-interacting system ($\lambda=1$) becomes
\begin{equation}
 	\label{Eq:BGE-total}
	\begin{split}
		E_{tot}^{\textrm{BGE}}&=\sum_{a}^{occ}\epsilon_a+\epsilon_{a}^{1st}(1)+\sum_{a<b}^{occ}e_{ab}^{\textrm{BGE}}[n](1)\\
		&=E_{tot}^{\textrm{EX}} + E_{c}^{\textrm{BGE}}[n](1)
	\end{split}
\end{equation}
where $E_{tot}^{\textrm{EX}}$ is the exact-exchange total energy in the KS-DFT framework. BGE is thus exact for one- and two-electron systems, but approximate for more electrons, because  interaction terms between three or more electrons are missing.

\begin{figure}[!bp]
    \pgfdeclaredecoration{complete sines}{initial}
    {
        \state{initial}[
            width=+0pt,
            next state=sine,
            persistent precomputation={\pgfmathsetmacro\matchinglength{
                \pgfdecoratedinputsegmentlength / int(\pgfdecoratedinputsegmentlength/\pgfdecorationsegmentlength)}
                \setlength{\pgfdecorationsegmentlength}{\matchinglength pt}
            }] {}
        \state{sine}[width=\pgfdecorationsegmentlength]{
            \pgfpathsine{\pgfpoint{0.25\pgfdecorationsegmentlength}{0.5\pgfdecorationsegmentamplitude}}
            \pgfpathcosine{\pgfpoint{0.25\pgfdecorationsegmentlength}{-0.5\pgfdecorationsegmentamplitude}}
            \pgfpathsine{\pgfpoint{0.25\pgfdecorationsegmentlength}{-0.5\pgfdecorationsegmentamplitude}}
            \pgfpathcosine{\pgfpoint{0.25\pgfdecorationsegmentlength}{0.5\pgfdecorationsegmentamplitude}}
    }
        \state{final}{}
    }
    \tikzset{wave/.style={decorate, decoration={complete sines, amplitude=2pt, segment length=3pt}}}
	\begin{tikzpicture}[scale=\myscale, decoration={markings, mark=at position 0.6 with {\arrow{latex}}}]
		\path (-0.2,2.0) coordinate (n1);
		\path (-0.2,0.5) coordinate (n2);
		\path (0.7,0.5) coordinate (p5-center);
		\path (0.2,0.0) coordinate (p4-a)
		      (0.2,1.0) coordinate (p5-a);
		\path (1.2,0.0) coordinate (p4-b)
		      (1.2,1.0) coordinate (p5-b);
		\path (1.85,0.5) coordinate (p02-b);
		\path (3.0,0.5) coordinate (p5-center-2);
		\path (2.5,0.0) coordinate (p4-a-2)
		      (2.5,1.0) coordinate (p5-a-2);
		\path (3.5,0.0) coordinate (p4-b-2)
		      (3.5,1.0) coordinate (p5-b-2);
		\path (2.81,0.70) coordinate (p6-a-2)
		      (3.20,0.70) coordinate (p6-b-2);
		\path (0.2,1.5) coordinate (p2-a)
		      (0.2,2.5) coordinate (p3-a);
		\path (3.5,0.0) coordinate (p6-b)
		      (3.5,1.0) coordinate (p7-b);
		\path (1.2,1.5) coordinate (p2-b)
		      (1.2,2.5) coordinate (p3-b);
		\path (1.85,2.0) coordinate (p01-b);
		\path (2.5,1.5) coordinate (p2-c)
		      (2.8,2.0) coordinate (p01-c)
		      (2.5,2.5) coordinate (p3-c);
		\path (3.1,0.0) coordinate (p4-d)
		      (3.5,1.0) coordinate (p5-d);
		\path (3.5,0.5) coordinate (p02-d)
		      (3.1,0.5) coordinate (p02-c);
		\path (3.5,1.5) coordinate (p2-d)
		      (3.2,2.0) coordinate (p01-d)
		      (3.5,2.5) coordinate (p3-d);
		\path (4.0,2.0) coordinate (p01-e)
		      (4.0,0.5) coordinate (p02-e);

		\node[left] at (n1) {\fontsize{\ptb}{1em}\selectfont $e_{ab}^{\textrm{BGE}} = $};
		\draw[line width=0.5] (p2-a) edge[bend left=60] node {\fontsize{\pta}{1em}\selectfont $\blacktriangledown$} (p3-a);
		\draw[line width=0.5] (p2-a) edge[bend right=60] node {\fontsize{\pta}{1em}\selectfont $\blacktriangle$} (p3-a);
		\draw[line width=0.5, wave] (p2-a) -- (p2-b);
		\draw[line width=0.5, wave] (p3-a) -- (p3-b);
		\draw[line width=0.5] (p2-b) edge[bend left=60] node {\fontsize{\pta}{1em}\selectfont $\blacktriangle$} (p3-b);
		\draw[line width=0.5] (p2-b) edge[bend right=60] node {\fontsize{\pta}{1em}\selectfont $\blacktriangledown$} (p3-b);
		\node at (p01-b) {\fontsize{\ptb}{1em}\selectfont $+$};
		\draw[line width=0.5] (p2-c) edge[bend left=60] node {\fontsize{\pta}{1em}\selectfont $\blacktriangledown$} (p3-c);
		\draw[line width=0.5] (p2-c) edge[bend right=60] node {\fontsize{\pta}{1em}\selectfont $\blacktriangle$} (p3-c);
		\draw[line width=0.5, wave] (p2-c) -- (p2-d);
		\draw[line width=0.5, wave] (p3-c) -- (p3-d);
		\draw[line width=0.5, wave] (p01-c) -- (p01-d);
		\draw[line width=0.5] (p2-d) edge[bend left=60] node {\fontsize{\pta}{1em}\selectfont $\blacktriangle$} (p3-d);
		\draw[line width=0.5] (p2-d) edge[bend right=60] node {\fontsize{\pta}{1em}\selectfont $\blacktriangledown$} (p3-d);
		\node[right] at (p01-e) {\fontsize{\ptb}{1em}\selectfont $+ \cdots$};

		\node[left] at (n2) {\fontsize{\ptb}{1em}\selectfont $+ $};
		\draw[line width=0.5] (p4-a) edge[bend left=60] node {\fontsize{\pta}{1em}\selectfont $\blacktriangledown$} (p5-a);
		\draw[line width=0.5, postaction={decorate}] (p4-a) -- (p5-center);
		\draw[line width=0.5] (p5-center) -- (p5-b);
		\draw[line width=0.5, wave] (p4-a) -- (p4-b);
		\draw[line width=0.5, wave] (p5-a) -- (p5-b);
		\draw[line width=0.5, postaction={decorate}] (p4-b) -- (p5-center);
		\draw[line width=0.5] (p5-a) -- (p5-center);
		\draw[line width=0.5] (p4-b) edge[bend right=60] node {\fontsize{\pta}{1em}\selectfont $\blacktriangle$} (p5-b);
		\node at (p02-b) {\fontsize{\ptb}{1em}\selectfont $+$};
		\draw[line width=0.5] (p4-a-2) edge[bend left=60] node {\fontsize{\pta}{1em}\selectfont $\blacktriangledown$} (p5-a-2);
		\draw[line width=0.5, postaction={decorate}] (p4-a-2) -- (p5-center-2);
		\draw[line width=0.5] (p5-center-2) -- (p5-b-2);
		\draw[line width=0.5, wave] (p4-a-2) -- (p4-b-2);
		\draw[line width=0.5, wave] (p5-a-2) -- (p5-b-2);
		\draw[line width=0.5, postaction={decorate}] (p4-b-2) -- (p5-center-2);
		\draw[line width=0.5] (p5-center-2) -- (p5-a-2);
		\draw[line width=0.5] (p4-b-2) edge[bend right=60] node {\fontsize{\pta}{1em}\selectfont $\blacktriangle$} (p5-b-2);
		\draw[line width=0.5, wave] (p6-a-2) -- (p6-b-2);

		\node[right] at (p02-e) {\fontsize{\ptb}{1em}\selectfont $+ \cdots$};
	\end{tikzpicture}
	\caption{\label{Fig:diagram}
		The Goldstone diagrams in the BGE are an infinite sequence of particle-particle ladder diagrams (pp-ladder)\cite{fetter:1996A}.
		The squiggly lines refer to the bare Coulomb interaction.
	}
\end{figure}
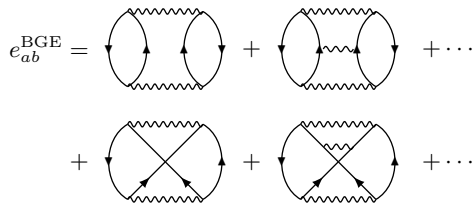

With the definition of $e_{ab}^{\textrm{BGE}}(\lambda)$, eq.~\ref{Eq:BGE-2} becomes
\begin{widetext}
\begin{equation}
	\label{Eq:Eab}
	\begin{split}
		e_{ab}^{\textrm{BGE}}(\lambda)
		=&\sum_{r<s}^{unocc}\frac{\lambda^2\left<\Phi_{ab}\left|\hat{H}_{1}(\lambda)\right|\Phi_{rs}\right>
		\left<\Phi_{rs}\left|\hat{H}_{1}(\lambda)\right|\Psi_{ab}\right>}
		{e_{ab}^{\textrm{BGE}}(\lambda)+e_{ab}^{1st}(\lambda)-\Delta\epsilon_{ab}^{rs}}\\
		=&\sum_{r<s}^{unocc}\frac{\left<\Phi_{ab}\left|\lambda\hat{H}_{1}(\lambda)\right|\Phi_{rs}\right>
		\left<\Phi_{rs}\left|\lambda\hat{H}_{1}(\lambda)\right|\Phi_{ab}\right>}
		{e_{ab}^{\textrm{BGE}}(\lambda)+e_{ab}^{1st}(\lambda)-\Delta\epsilon_{ab}^{rs}}\\
		&+\sum_{r<s}^{unocc}\sum_{p<q}^{unocc}
		\frac{\left<\Phi_{ab}\left|\lambda\hat{H}_{1}(\lambda)\right|\Phi_{rs}\right>
		\left<\Phi_{rs}\left|\lambda\hat{H}_{1}(\lambda)\right|\Phi_{pq}\right>\left<\Phi_{pq}\left|\lambda
		\hat{H}_{1}(\lambda)\right|\Psi_{ab}\right>}
		{(e_{ab}^{\textrm{BGE}}(\lambda)+e_{ab}^{1st}(\lambda)-\Delta\epsilon_{ab}^{rs})
		(e_{ab}^{\textrm{BGE}}(\lambda)+e_{ab}^{1st}(\lambda)-\Delta\epsilon_{ab}^{pq})}\\
		=&\cdots
	\end{split}
\end{equation}
\end{widetext}
where $\Delta\epsilon_{ab}^{rs}=\epsilon_{r}+\epsilon_{s}-\epsilon_{a}-\epsilon_{b}$. The second term on the third line of 
eq.~\ref{Eq:Eab} emerges when we replace $\Psi_{ab}$ by eq.~\ref{Eq:BGE-1}.
This expansion reveals that the BGE correlation energy contains an infinite summation of a sequence of Goldstone
pair diagrams\cite{fetter:1996A,szabo:1996A}. Therefore, this sequence of Goldstone diagrams contains only two hole lines, representing the 
electron pair $ab$, and the infinite summation goes through all the ladder diagrams over two particle lines (see fig.~\ref{Fig:diagram}).
In other words, the intermediate pairs always propagate as electrons\cite{fetter:1996A}. Conversely,
eq.~\ref{Eq:Eab} has to be solved self-consistently, as the electron-pair energy $E_{ab}$ depends on itself.

As alluded to before, the BGE accounts for the correlation of two electrons and thus provides the exact solution for one- and 
two-electron systems. However, eq.~\ref{Eq:Eab} is not exact, because we omitted the single excitations at the very beginning. 
Nonetheless,  our approximation  should still be able to capture the subtle (near)-degeneracy static correlation effects at 
H$_2$ dissociation and eliminate the one-electron ``self-correlation'' error as does the configuration-interaction method 
with double excitations (CID) or the coupled-cluster doubles (CCD) method. Nesbet\cite{nesbet:1971A} has demonstrated that 
BGE is equivalent to the so-called independent electron-pair approximation (IEPA) in quantum chemistry\cite{szabo:1996A}, 
which can be considered as an intermediate approximation between CID and MP2. However, we prefer to keep the BGE acronym 
(eq.~\ref{Eq:Eab}) as it is more compact and easily linked to an expansion of Goldstone diagrams\cite{szabo:1996A,brueckner:1955A,goldstone:1957A} 
which is helpful for further discussions (fig.~\ref{Fig:diagram}).

\section{\label{Sec:BGE2} The second-order BGE approximation}
\subsection{Derivation of BGE2}
The BGE electron-pair correlation energy $e_{ab}^{\textrm{BGE}}$ in eq.~\ref{Eq:Eab} contains an infinite sequence 
of particle-particle ladder diagrams (pp-ladder resummation), as shown in fig.~\ref{Fig:diagram}, because the BGE
wave function $\Psi_{ab}$ also appears on the right-hand side of the equation. In addition,  eq.~\ref{Eq:Eab} should 
be solved iteratively as $e_{ab}^{\textrm{BGE}}$ appears on both  sides of the equation ($e_{ab}$-coupling effect). 
These two mechanisms cooperate to deliver an accurate description of exchange and correlation in one- and two-electron 
systems. It has been argued that an explicit (or implicit) resummation of a selected series of diagrams (e.g., the
pp-ladder resummation shown in fig.\ \ref{Fig:diagram}) is necessary to remove the divergence at degeneracies of any
finite-order perturbation theory \cite{rinke:2013B,perdew:2000A,burke:2005A,kresse:2008A,olsen:2014A}. However, in this 
work we will show that the same effect can be achieved by the $e_{ab}$-coupling effect at finite orders of perturbation
theory. This allows us to terminate the BGE expansion at the second order, as long as we retain the $e_{ab}$-coupling 
effect. This second-order BGE  (BGE2) is the second approximation we make:
\begin{equation}
	\label{Eq:BGE2}
	\begin{split}
		e_{ab}^{\textrm{BGE2}}(\lambda)
		=&\sum_{r<s}^{unocc}\frac{\lambda^2\left|\left<\Phi_{ab}\left|\hat{H}_{1}(\lambda)\right|\Phi_{rs}\right>\right|^2}
		{e_{ab}^{\textrm{BGE2}}(\lambda)+e_{ab}^{1st}(\lambda)-\Delta\epsilon_{ab}^{rs}}\\
		=&\sum_{r<s}^{unocc}\frac{\lambda^2\left|\left<\Phi_{ab}\left|\hat{V}_{ee}\right|\Phi_{rs}\right>\right|^2}
		{e_{ab}^{\textrm{BGE2}}(\lambda)+e_{ab}^{1st}(\lambda)-\Delta\epsilon_{ab}^{rs}}\\
		=&\sum_{r<s}^{unocc}\frac{\lambda^2\left|\left<\phi_{a}\phi_{b}||\phi_{r}\phi_{s}\right>\right|^2}
		{e_{ab}^{\textrm{BGE2}}(\lambda)+e_{ab}^{1st}(\lambda)-\Delta\epsilon_{ab}^{rs}}.
	\end{split}
\end{equation}
Here we have utilized the fact that both $\hat{v}_{ab}^{\textrm{MF}}$ and $\hat{v}_{\lambda}$ are one-electron operators, 
which do not contribute to the expectation value between the ground state and a double excitation. 

We will show in Sec.~\ref{SSec:H2min} that BGE2 only dissociates H$_2$ in a minimal basis correctly, if we remove $e_{ab}^{1st}$ from
the denominator. So in the following we drop $e_{ab}^{1st}$. This is our final approximation:
\begin{equation}
	\label{Eq:BGE2a}
	\begin{split}
		e_{ab}^{\textrm{BGE2}}(\lambda)
		\approx &\sum_{r<s}^{unocc}\frac{\lambda^2\left|\left<\phi_{a}\phi_{b}||\phi_{r}\phi_{s}\right>\right|^2}
		{e_{ab}^{\textrm{BGE2}}(\lambda)-\Delta\epsilon_{ab}^{rs}}.
	\end{split}
\end{equation}
$e_{ab}^{\textrm{BGE2}}(\lambda)$ now appears as a simple sum-over-state formula that is similar to standard PT2 and thus 
exhibits the same computational scaling. We will again need to sum all electron pairs to obtain the full BGE2 correlation 
energy $E_{c}^{\textrm{BGE2}}(\lambda)$ for systems with more than two electrons
\begin{equation}
	\label{Eq:BGE2total}
	E_{c}^{\textrm{BGE2}}(\lambda)=\sum_{a<b}^{occ}e_{ab}^{\textrm{BGE2}}(\lambda).
\end{equation}
A distinct advantage of eq.~\ref{Eq:BGE2a} is that the dependence on the  coupling constant $\lambda$ is now simple and 
well-defined. Following eq.~\ref{Eq:AC} we can easily obtain the BGE2 electron-pair correlation potential $v_{ab}^{\textrm{BGE2}}(\lambda)$ as
\begin{equation}
	\label{Eq:BGE2Vab}
	\begin{split}
		v_{ab}^{\textrm{BGE2}}(\lambda)=&\sum_{r<s}^{unocc}\frac{2\lambda\left|\left<\phi_{a}\phi_{b}||\phi_{r}\phi_{s}\right>\right|^2}
		{e_{ab}^{\textrm{BGE2}}(\lambda)-\Delta\epsilon_{ab}^{rs}}\times\\
		&\left(1+\frac{1}{2}\frac{\lambda v_{ab}^{\textrm{BGE2}}(\lambda)}{e_{ab}^{\textrm{BGE2}}(\lambda)-\Delta\epsilon_{ab}^{rs}}\right)
	\end{split}
\end{equation}
and then for a many electron system
\begin{equation}
	V_{c}^{\textrm{BGE2}}(\lambda)=\sum_{a<b}^{occ}v_{ab}^{\textrm{BGE2}}(\lambda).
\end{equation}
The BGE2 correlation potential also has to be solved iteratively, which  prevents us from making further analytical manipulations. However, for H$_2$ in a minimal basis, the BGE2 correlation energy and potential can be solved analytically, which will give us more insight into BGE2. This  will be discussed later in Sec.~\ref{SSec:H2min}.

At this point, we will recap the approximations made in the derivation of the BGE2 correlation functional:
\begin{enumerate}
	\item From the outset we chose a pair theory. The full BGE (eqs.~\ref{Eq:pair0} and \ref{Eq:pair}) explicitly treats interactions in one electron pair and is exact for one- and two-electron systems. For more than two electrons, the interaction from other 
		electrons can be taken into account in a mean field fashion (eq.~\ref{Eq:MF}). Then the correlation energy sums 
		up the correlations of all electron pairs (eqs.~\ref{Eq:BGE-Eab} and \ref{Eq:BGE2total}).

	\item The BGE can be solved by means of Green's functions (eq.~\ref{Eq:BGE-GF}). Here we omit
		the single excitation contribution in the construction of the non-interacting Green's function $G_{0}$ (eq.~\ref{Eq:GF0}). We argue
		that this approximation is justified and captures the subtle exchange-correlation effects in one- and two-electron systems as does
		CID or CCD.
	
	\item The next approximation is to terminate the BGE expansion at the second order (eq.~\ref{Eq:BGE2}). This implies that we remove
		the infinite summation of particle-particle ladder diagrams (fig.~\ref{Fig:diagram}). The resulting BGE2 approximation 
		still retains the $e_{ab}$-coupling effect. We will show later that the $e_{ab}$-coupling at second-order in perturbation
		theory is sufficient to capture correlations that emerge from (near)-degeneracies and that higher-order connected Goldstone 
		diagrams are then not needed. 
		However, if we are far from degeneracies (e.g., H$_2$ in the middle of dissociation) BGE2 alone is 
		not sufficient and higher order diagrams would be required.

	\item The final approximation (eq.~\ref{Eq:BGE2a}) removes the first-order perturbation term $e_{ab}^{1st}$ from the denominator of 
		eq.~\ref{Eq:BGE2}. On the one hand, this omission removes the difficulty of having to consider the unknown density adaptive operator $\hat{v}_{\lambda}$ explicitly 
		along the AC path (see eqs.\ \ref{Eq:CS} and \ref{Eq:BGE-1st}). On the other hand, we will show in Sec.~\ref{SSec:H2min} that together 
		with the other two approximations, this approximation is necessary to deliver an accurate description of H$_2$ dissociation in a minimal basis.
\end{enumerate}

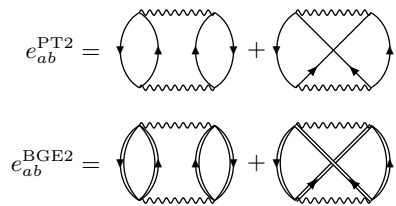
\begin{figure}[!bp]
    \pgfdeclaredecoration{complete sines}{initial}
    {
        \state{initial}[
            width=+0pt,
            next state=sine,
            persistent precomputation={\pgfmathsetmacro\matchinglength{
                \pgfdecoratedinputsegmentlength / int(\pgfdecoratedinputsegmentlength/\pgfdecorationsegmentlength)}
                \setlength{\pgfdecorationsegmentlength}{\matchinglength pt}
            }] {}
        \state{sine}[width=\pgfdecorationsegmentlength]{
            \pgfpathsine{\pgfpoint{0.25\pgfdecorationsegmentlength}{0.5\pgfdecorationsegmentamplitude}}
            \pgfpathcosine{\pgfpoint{0.25\pgfdecorationsegmentlength}{-0.5\pgfdecorationsegmentamplitude}}
            \pgfpathsine{\pgfpoint{0.25\pgfdecorationsegmentlength}{-0.5\pgfdecorationsegmentamplitude}}
            \pgfpathcosine{\pgfpoint{0.25\pgfdecorationsegmentlength}{0.5\pgfdecorationsegmentamplitude}}
    }
        \state{final}{}
    }
    \tikzset{wave/.style={decorate, decoration={complete sines, amplitude=2pt, segment length=3pt}}}
	\begin{tikzpicture}[scale=\myscale]
		\path (-0.2,2.0) coordinate (n1);
		\path (2.00,2.0) coordinate (n2);
		\path (2.78,2.0) coordinate (p5-center);
		\path (2.28,1.5) coordinate (p4-a)
		      (2.28,2.5) coordinate (p5-a);
		\path (3.28,1.5) coordinate (p4-b)
		      (3.28,2.5) coordinate (p5-b);
		\path (0.2,1.5) coordinate (p2-a)
		      (0.2,2.5) coordinate (p3-a);
		\path (1.2,1.5) coordinate (p2-b)
		      (1.2,2.5) coordinate (p3-b);
		\path (-0.2,0.5) coordinate (b1);
		\path (2.00,0.5) coordinate (b2);
		\path (2.78,0.5) coordinate (b5-center);
		\path (2.28,0.0) coordinate (b4-a)
		      (2.28,1.0) coordinate (b5-a);
		\path (3.28,0.0) coordinate (b4-b)
		      (3.28,1.0) coordinate (b5-b);
		\path (0.2,0.0) coordinate (b2-a)
		      (0.2,1.0) coordinate (b3-a);
		\path (1.2,0.0) coordinate (b2-b)
		      (1.2,1.0) coordinate (b3-b);
		\node[left] at (n1) {\fontsize{\ptb}{1em}\selectfont $e_{ab}^{\textrm{PT2}} = $};
		\draw[line width=0.5] (p2-a) edge[bend left=60] node {\fontsize{\pta}{1em}\selectfont $\blacktriangledown$} (p3-a);
		\draw[line width=0.5] (p2-a) edge[bend right=60] node {\fontsize{\pta}{1em}\selectfont $\blacktriangle$} (p3-a);
		\draw[line width=0.5, wave] (p2-a) -- (p2-b);
		\draw[line width=0.5, wave] (p3-a) -- (p3-b);
		\draw[line width=0.5] (p2-b) edge[bend left=60] node {\fontsize{\pta}{1em}\selectfont $\blacktriangle$} (p3-b);
		\draw[line width=0.5] (p2-b) edge[bend right=60] node {\fontsize{\pta}{1em}\selectfont $\blacktriangledown$} (p3-b);
		\node[left] at (n2) {\fontsize{\ptb}{1em}\selectfont $+ $};
		\draw[line width=0.5] (p4-a) edge[bend left=60] node {\fontsize{\pta}{1em}\selectfont $\blacktriangledown$} (p5-a);
		\draw[line width=0.5, decoration={markings, mark=at position 0.6 with {\arrow{latex}}}, postaction={decorate}] (p4-a) -- (p5-center);
		\draw[line width=0.5] (p5-center) -- (p5-b);
		\draw[line width=0.5, wave] (p4-a) -- (p4-b);
		\draw[line width=0.5, wave] (p5-a) -- (p5-b);
		\draw[line width=0.5, decoration={markings, mark=at position 0.6 with {\arrow{latex}}}, postaction={decorate}] (p4-b) -- (p5-center);
		\draw[line width=0.5] (p5-a) -- (p5-center);
		\draw[line width=0.5] (p4-b) edge[bend right=60] node {\fontsize{\pta}{1em}\selectfont $\blacktriangle$} (p5-b);
		
		\node[left] at (b1) {\fontsize{\ptb}{1em}\selectfont $e_{ab}^{\textrm{BGE2}} = $};
		\draw[line width=0.5] (b2-a) edge[bend left=60] node {\fontsize{\pta}{1em}\selectfont $\blacktriangledown$} (b3-a);
		\draw[line width=0.5] (b2-a) edge[bend left=45] (b3-a);
		\draw[line width=0.5] (b2-a) edge[bend right=60] node {\fontsize{\pta}{1em}\selectfont $\blacktriangle$} (b3-a);
		\draw[line width=0.5] (b2-a) edge[bend right=45] (b3-a);
		\draw[line width=0.5, wave] (b2-a) -- (b2-b);
		\draw[line width=0.5, wave] (b3-a) -- (b3-b);
		\draw[line width=0.5] (b2-b) edge[bend left=60] node {\fontsize{\pta}{1em}\selectfont $\blacktriangle$} (b3-b);
		\draw[line width=0.5] (b2-b) edge[bend left=45] (b3-b);
		\draw[line width=0.5] (b2-b) edge[bend right=60] node {\fontsize{\pta}{1em}\selectfont $\blacktriangledown$} (b3-b);
		\draw[line width=0.5] (b2-b) edge[bend right=45] (b3-b);
		\node[left] at (b2) {\fontsize{\ptb}{1em}\selectfont $+ $};
		\draw[line width=0.5] (b4-a) edge[bend left=60] node {\fontsize{\pta}{1em}\selectfont $\blacktriangledown$} (b5-a);
		\draw[line width=0.5] (b4-a) edge[bend left=45] (b5-a);
		\draw[line width=0.5, decoration={markings, mark=at position 0.6 with {\arrow[scale=0.6]{latex}}}, postaction={decorate},double] (b4-a) -- (b5-center);
		\draw[line width=0.5,double] (b5-center) -- (b5-b);
		\draw[line width=0.5, wave] (b4-a) -- (b4-b);
		\draw[line width=0.5, wave] (b5-a) -- (b5-b);
		\draw[line width=0.5, decoration={markings, mark=at position 0.6 with {\arrow[scale=0.6]{latex}}}, postaction={decorate},double] (b4-b) -- (b5-center);
		\draw[line width=0.5,double] (b5-a) -- (b5-center);
		\draw[line width=0.5] (b4-b) edge[bend right=60] node {\fontsize{\pta}{1em}\selectfont $\blacktriangle$} (b5-b);
		\draw[line width=0.5] (b4-b) edge[bend right=45] (b5-b);
	\end{tikzpicture}
	\caption{\label{Fig:diagram-BGE2}
		The diagramatic representation of PT2 and BGE2. Double lines in the BGE2 diagram represent a correction to the double 
		excitation energies due to the $e_{ab}$-coupling effect, which should be solved iteratively. The squiggly lines refer to the bare Coulomb interaction.
	}
\end{figure}

\subsection{Analysis of the $e_{ab}$-coupling effect in BGE2}
Comparing to the standard PT2 expression, the only difference in the BGE2 correlation expression (eqs.~\ref{Eq:BGE2a} and~\ref{Eq:BGE2total}) is that 
the BGE2 electron-pair correlation $e_{ab}^{\textrm{BGE2}}(\lambda)$ itself appears in the denominator. 
$e_{ab}^{\textrm{BGE2}}(\lambda)$ should acquire a finite value to prevent the numerical divergence for small-gap systems
where $\Delta\epsilon_{ab}^{rs}\rightarrow 0$. 
To distinguish BGE2 from PT2, modified particle and hole lines (double line) are introduced in fig.~\ref{Fig:diagram-BGE2} to represent the $e_{ab}$-coupling
effect in the BGE2 method which corrects the double-excitation energies $\Delta\epsilon_{ab}^{rs}$ and should be solved iteratively. 
We will demonstrate the accuracy of the $e_{ab}$-coupling effect later both numerically (Sec. \ref{SSec:Num}) and analytically
(Sec. \ref{SSec:H2min}). In this section, we will provide a many-body perturbation theory perspective of the $e_{ab}$-coupling effect.

In quantum chemistry, the configuration interaction equation with singles and doubles (CISD) is usually solved with iterative
techniques\cite{nesbet:1965A}, to avoid a direct diagonalization of the large Hamiltonian matrices in  configuration space.
Pople {\it et. al.} \cite{pople:1977A} demonstrated that such iterative algorithms are more than just a technical trick. From a  
many-body perturbation theory viewpoint, each iteration introduces higher order terms. For example, after the second iteration, 
the second- and third-order terms emerge in the CISD energy
expression but with a scaled weight\cite{pople:1977A,shavitt:1973A}. This would also be true for the $e_{ab}$-coupling effect in the
complete BGE expansion (eq.~\ref{Eq:Eab}). But can we write down a perturbative expansion for
the second-order BGE expression (eq.~\ref{Eq:BGE2a} and fig.~\ref{Fig:diagram-BGE2})? In other words,  does the $e_{ab}$-coupling effect introduce higher order perturbation terms during the iterative procedure?
In this section, we will answer these questions step-by-step.

To demonstrate the behavior of BGE2 for (near)-degeneracies, we analyse its limit as $\Delta\epsilon_{ab}^{rs}$ goes to zero. 
To do so we need to introduce a level-shift ($L$) into the expression of  the BGE2 correlation energy (eq.~\ref{Eq:BGE2a})
\begin{equation}
	\label{Eq:BGE2am}
	\begin{split}
		e_{ab}^{\textrm{BGE2}}(\lambda)
		= &\sum_{r<s}^{unocc}\frac{\lambda^2\left|\left<\phi_{a}\phi_{b}||\phi_{r}\phi_{s}\right>\right|^2}
		{e_{ab}^{\textrm{BGE2}}(\lambda)+L-(\Delta\epsilon_{ab}^{rs}+L)}\\
		= &-\sum_{r<s}^{unocc}\frac{\lambda^2\left|\left<\phi_{a}\phi_{b}||\phi_{r}\phi_{s}\right>\right|^2}
		{(\Delta\epsilon_{ab}^{rs}+L)}\left( 1-x(L) \right)^{-1}\\
	\end{split}
\end{equation}
where
\begin{equation}
		x(L) = \frac{e_{ab}^{\textrm{BGE2}}(\lambda)+L}{\Delta\epsilon_{ab}^{rs}+L}.
\end{equation}
To be able to expand eq.~\ref{Eq:BGE2am} into a geometric series we require
\begin{equation}
	-1<x(L)<1 \quad .
\end{equation}
This leads to the following constraint for $L$
\begin{equation}
	\label{Eq:level-shift}
	L>\max\left( 0,-\frac{1}{2}\left( \Delta\epsilon_{ab}^{rs}+e_{ab}^{\textrm{BGE2}}(\lambda) \right) \right).
\end{equation}
By definition we have $e_{ab}^{\textrm{BGE2}}\le 0$ and $\Delta\epsilon_{ab}^{rs}\ge 0$. In addition, $\Delta\epsilon_{ab}^{rs}> |e_{ab}^{\textrm{BGE2}}(\lambda)|$
holds for insulators, most semi-conductors and even for most of the double excitations in small-gap systems (excluding cases where $ab$ refers to the highest 
occupied molecular orbital (HOMO)). We can then always choose $L$=0. 

However, we cannot take $L$=0 when the energy gap of double excitations from the HOMO to the LUMO ($\Delta\epsilon_{ab}^{rs}$) tends to zero and 
$\Delta\epsilon_{ab}^{rs}\le |e_{ab}^{\textrm{BGE2}}(\lambda)|$. Then only a positive level shift 
$L>-\frac{1}{2}(\Delta\epsilon_{ab}^{rs}+e_{ab}^{\textrm{BGE2}}(\lambda))$ will guarantee a convergent geometric expansion. 

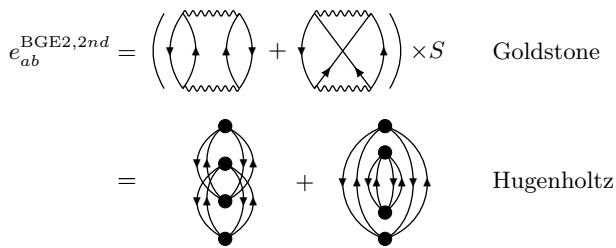
\begin{figure}[!tbp]
    \pgfdeclaredecoration{complete sines}{initial}
    {
        \state{initial}[
            width=+0pt,
            next state=sine,
            persistent precomputation={\pgfmathsetmacro\matchinglength{
                \pgfdecoratedinputsegmentlength / int(\pgfdecoratedinputsegmentlength/\pgfdecorationsegmentlength)}
                \setlength{\pgfdecorationsegmentlength}{\matchinglength pt}
            }] {}
        \state{sine}[width=\pgfdecorationsegmentlength]{
            \pgfpathsine{\pgfpoint{0.25\pgfdecorationsegmentlength}{0.5\pgfdecorationsegmentamplitude}}
            \pgfpathcosine{\pgfpoint{0.25\pgfdecorationsegmentlength}{-0.5\pgfdecorationsegmentamplitude}}
            \pgfpathsine{\pgfpoint{0.25\pgfdecorationsegmentlength}{-0.5\pgfdecorationsegmentamplitude}}
            \pgfpathcosine{\pgfpoint{0.25\pgfdecorationsegmentlength}{0.5\pgfdecorationsegmentamplitude}}
    }
        \state{final}{}
    }

    \tikzset{round left paren/.style={out=120,in=-120}}
    \tikzset{round right paren/.style={out=60,in=-60}}
    
    \tikzset{wave/.style={decorate, decoration={complete sines, amplitude=2pt, segment length=3pt}}}
    
	\begin{tikzpicture}[scale=\myscaletwo]

		\path (0.0,2.0) coordinate (n1)
		      (4.25,-1.5) coordinate (n2)
		      (0.0,-1.5) coordinate (n3)
		      (4.0,2.0) coordinate (n4)
		      (7.5,2.0) coordinate (n5);
		\path (0.6,-2.5) coordinate (p0-a)
		      (0.6,-0.5) coordinate (p1-a);
		\path (3.4,-2.5) coordinate (p0-b)
		      (3.4,-0.5) coordinate (p1-b);

		\path (5.1,-2.5) coordinate (p0-c)
		      (5.1,-0.5) coordinate (p1-c);
		\path (7.9,-2.5) coordinate (p0-d)
		      (7.9,-0.5) coordinate (p1-d);

		\path (1.5,-2.0) coordinate (p2-a)
		      (1.5,-1.0) coordinate (p3-a);
		\path (2.5,-2.0) coordinate (p2-b)
		      (2.5,-1.0) coordinate (p3-b);

		\path (6.0,-1.5) coordinate (p2-c)
		      (6.0, 0.5) coordinate (p3-c);
		\path (7.0,-1.5) coordinate (p2-d)
		      (7.0, 0.5) coordinate (p3-d);
		\path (0.5,1.0) coordinate (b1)
		      (0.5,3.0) coordinate (b2)
		      (6.5,1.0) coordinate (b3)
		      (6.5,3.0) coordinate (b4);
		\path (1.0,1.0) coordinate (p4-a)
		      (1.0,3.0) coordinate (p5-a);
		\path (2.5,1.0) coordinate (p4-b)
		      (2.5,3.0) coordinate (p5-b);
		\path (5.25,2.0) coordinate (e5-center);
		\path (4.5,1.0) coordinate (e4-a)
		      (4.5,3.0) coordinate (e5-a);
		\path (6.0,1.0) coordinate (e4-b)
		      (6.0,3.0) coordinate (e5-b);
		
		\path (9.0,2.0) coordinate (note1)
		      (9.0,-1.5) coordinate (note2);

		\node[left] at (n1) {\fontsize{\ptb}{1em}\selectfont $e_{ab}^{\textrm{BGE2},2nd} = $};
		\draw[line width=0.5] (b1) to [round left paren ] (b2);
		\draw[line width=0.5] (b3) to [round right paren] (b4);

		\node[left] at (n3) {\fontsize{\ptb}{1em}\selectfont $ = $};
		\draw[line width=0.5] (p4-a) edge[bend left=36] node {\fontsize{\pta}{1em}\selectfont $\blacktriangledown$} (p5-a);
		\draw[line width=0.5] (p4-a) edge[bend right=36] node {\fontsize{\pta}{1em}\selectfont $\blacktriangle$} (p5-a);
		\draw[line width=0.5] (p4-b) edge[bend left=36] node {\fontsize{\pta}{1em}\selectfont $\blacktriangle$} (p5-b);
		\draw[line width=0.5] (p4-b) edge[bend right=36] node {\fontsize{\pta}{1em}\selectfont $\blacktriangledown$} (p5-b);
		\draw[line width=0.5, wave] (p4-a) -- (p4-b);
		\draw[line width=0.5, wave] (p5-a) -- (p5-b);

		\node[left] at (n4) {\fontsize{\ptb}{1em}\selectfont $+ $};
		\draw[line width=0.5] (e4-a) edge[bend left=36] node {\fontsize{\pta}{1em}\selectfont $\blacktriangledown$} (e5-a);
		\draw[line width=0.5, decoration={markings, mark=at position 0.6 with {\arrow{latex}}}, postaction={decorate}] (e4-a) -- (e5-center);
		\draw[line width=0.5] (e5-center) -- (e5-b);
		\draw[line width=0.5, wave] (e4-a) -- (e4-b);
		\draw[line width=0.5, wave] (e5-a) -- (e5-b);
		\draw[line width=0.5, decoration={markings, mark=at position 0.6 with {\arrow{latex}}}, postaction={decorate}] (e4-b) -- (e5-center);
		\draw[line width=0.5] (e5-a) -- (e5-center);
		\draw[line width=0.5] (e4-b) edge[bend right=36] node {\fontsize{\pta}{1em}\selectfont $\blacktriangle$} (e5-b);
		
		\node at (n5) {\fontsize{\ptb}{1em}\selectfont $\times S$};
		
		\path (2.125,-2.0) coordinate (p0-a)
		      (2.125,-0.0) coordinate (p1-a);
		\path (2.125,-3.0) coordinate (p0-b)
		      (2.125,-1.0) coordinate (p1-b);
		\path (4.25,-1.5) coordinate (n2)
		      (0.0,-1.5) coordinate (n3);
		\path (6.375,-3.0) coordinate (p2-a)
		      (6.375,-0.0) coordinate (p3-a);
		\path (6.375,-2.3) coordinate (p2-b)
		      (6.375,-0.7) coordinate (p3-b);
		
		\node[left] at (n3) {\fontsize{\ptb}{1em}\selectfont $ = $};
		
		\draw[fill=black] (p0-a) circle [radius=5pt];
		\draw[fill=black] (p1-a) circle [radius=5pt];
		\draw[fill=black] (p0-b) circle [radius=5pt];
		\draw[fill=black] (p1-b) circle [radius=5pt];
		\draw[line width=0.5] (p0-a) .. controls ++(-1.0, 0.5) and ++(-1.0,-0.5) .. node {\fontsize{\pta}{1em}\selectfont $\blacktriangledown$} (p1-a);
                \draw[line width=0.5] (p0-a) .. controls ++(1.0, 0.5) and ++(1.0,-0.5) .. node {\fontsize{\pta}{1em}\selectfont $\blacktriangle$} (p1-a);
                \draw[line width=0.5] (p0-a) .. controls ++(-0.6, 0.4) and ++(-0.7,-0.4) .. node {\fontsize{\pta}{1em}\selectfont $\blacktriangle$} (p1-a);
                \draw[line width=0.5] (p0-a) .. controls ++(0.6, 0.4) and ++(0.7,-0.4) .. node {\fontsize{\pta}{1em}\selectfont $\blacktriangledown$} (p1-a);
                \draw[line width=0.5] (p0-b) .. controls ++(-1.0, 0.5) and ++(-1.0,-0.5) .. node {\fontsize{\pta}{1em}\selectfont $\blacktriangledown$} (p1-b);
                \draw[line width=0.5] (p0-b) .. controls ++(1.0, 0.5) and ++(1.0,-0.5) .. node {\fontsize{\pta}{1em}\selectfont $\blacktriangle$} (p1-b);
                \draw[line width=0.5] (p0-b) .. controls ++(-0.6, 0.4) and ++(-0.7,-0.4) .. node {\fontsize{\pta}{1em}\selectfont $\blacktriangle$} (p1-b);
                \draw[line width=0.5] (p0-b) .. controls ++(0.6, 0.4) and ++(0.7,-0.4) .. node {\fontsize{\pta}{1em}\selectfont $\blacktriangledown$} (p1-b);
		
		\draw[fill=black] (p2-a) circle [radius=5pt];
		\draw[fill=black] (p3-a) circle [radius=5pt];
		\draw[fill=black] (p2-b) circle [radius=5pt];
		\draw[fill=black] (p3-b) circle [radius=5pt];
                \draw[line width=0.5] (p2-a) .. controls ++(-1.5, 0.5) and ++(-1.5,-0.5) .. node {\fontsize{\pta}{1em}\selectfont $\blacktriangledown$} (p3-a);
                \draw[line width=0.5] (p2-a) .. controls ++(1.5, 0.5) and ++(1.5,-0.5) .. node {\fontsize{\pta}{1em}\selectfont $\blacktriangle$} (p3-a);
                \draw[line width=0.5] (p2-a) .. controls ++(-1.0, 0.5) and ++(-1.0,-0.5) .. node {\fontsize{\pta}{1em}\selectfont $\blacktriangle$} (p3-a);
                \draw[line width=0.5] (p2-a) .. controls ++(1.0, 0.5) and ++(1.0,-0.5) .. node {\fontsize{\pta}{1em}\selectfont $\blacktriangledown$} (p3-a);
                
                \draw[line width=0.5] (p2-b) .. controls ++(-0.6, 0.5) and ++(-0.6,-0.5) .. node {\fontsize{\pta}{1em}\selectfont $\blacktriangledown$} (p3-b);
                \draw[line width=0.5] (p2-b) .. controls ++(0.6, 0.5) and ++(0.6,-0.5) .. node {\fontsize{\pta}{1em}\selectfont $\blacktriangle$} (p3-b);
                \draw[line width=0.5] (p2-b) .. controls ++(-0.3, 0.3) and ++(-0.3,-0.3) .. node {\fontsize{\pta}{1em}\selectfont $\blacktriangle$} (p3-b);
                \draw[line width=0.5] (p2-b) .. controls ++(0.3, 0.3) and ++(0.3,-0.3) .. node {\fontsize{\pta}{1em}\selectfont $\blacktriangledown$} (p3-b);

		\node at (n2) {\fontsize{\ptb}{1em}\selectfont $+$};

		\node[right] at (note1) {\fontsize{\ptb}{1em}\selectfont Goldstone};
		\node[right] at (note2) {\fontsize{\ptb}{1em}\selectfont Hugenholtz};
	\end{tikzpicture}
	\caption{\label{Fig:disconnected-4th-diagram}
	The BGE2 diagrams at second order are rescaled PT2-like diagrams represented by two disconnected Hugenholtz diagrams
	with quadrupole excitations, which can be expanded into 32 disconnected Goldstone diagrams~\cite{fetter:1996A, brian:1975A}.
	}
\end{figure}

We will discuss the positive level shift later and first analyse the $L$=0 case. Expanding eq.~\ref{Eq:BGE2am} into a geometric series yields
\begin{equation}
	\label{Eq:BGE2expansion}
	\begin{split}
		e_{ab}^{\textrm{BGE2}}(\lambda)
		= &-\sum_{n=0}^{\infty}\sum_{r<s}^{unocc}\frac{\lambda^2\left|\left<\phi_{a}\phi_{b}||\phi_{r}\phi_{s}\right>\right|^2}
		{(\Delta\epsilon_{ab}^{rs}+L)}x(L)^n ,
	\end{split}
\end{equation}
which for $L=0$ becomes
\begin{equation}
	\label{Eq:BGE2ame}
	\begin{split}
		e_{ab}^{\textrm{BGE2}}(\lambda)
		= &-\sum_{n=0}^{\infty}\sum_{r<s}^{unocc}\frac{\lambda^2\left|\left<\phi_{a}\phi_{b}||\phi_{r}\phi_{s}\right>\right|^2}
		{(\Delta\epsilon_{ab}^{rs})^{n+1}}e_{ab}^{\textrm{BGE2}}(\lambda)^n .
	\end{split}
\end{equation}
This expression allows us to analyze the BGE2 correlation energy from a many-body perspective by iterating the right-hand side.
Here, we examine the simplest two terms. The first term in the BGE2 expansion ($n=0$) is nothing but the standard PT2 correlation energy 
\begin{equation}
	\label{Eq:BGE2-1st}
	e_{ab}^{\textrm{BGE2},1st}(\lambda)=
		-\sum_{r<s}^{unocc}\frac{\lambda^2\left|\left<\phi_{a}\phi_{b}||\phi_{r}\phi_{s}\right>\right|^2}
		{\Delta\epsilon_{ab}^{rs}} .
\end{equation}
The Goldstone diagrams of the PT2 correlation are shown in fig.~\ref{Fig:diagram-BGE2}).
The second term becomes
\begin{equation}
	\label{Eq:BGE2-2nd}
	\begin{split}
		e_{ab}^{\textrm{BGE2},2nd}(\lambda)=
			&\sum_{r<s}^{unocc}\frac{\lambda^4\left|\left<\phi_{a}\phi_{b}||\phi_{r}\phi_{s}\right>\right|^2}
			{\Delta\epsilon_{ab}^{rs}}\times S\\
	\end{split}
\end{equation}
where $S$ is the normalization of the first-order perturbative wave-function $\Phi_{ab}^{1st}$ of the electron-pair $ab$
\begin{equation}
	\begin{split}
		&\left|\Phi_{ab}^{1st}\right>=\sum_{r<s}\frac{\left<\phi_{a}\phi_{b}||\phi_{r}\phi_{s}\right>}{\Delta\epsilon_{ab}^{rs}}\left|\Phi_{rs}\right> \\
		S=&\sum_{r<s}^{unocc}\frac{\left|\left<\phi_{a}\phi_{b}||\phi_{r}\phi_{s}\right>\right|^2}
		{\left(\Delta\epsilon_{ab}^{rs}\right)^2}
		=\left<\Phi_{ab}^{1st}|\Phi_{ab}^{1st}\right>
	\end{split}
\end{equation}
$e_{ab}^{\textrm{BGE2},2nd}$ is a fourth-order perturbation in terms of the coupling constant $\lambda$. 
This expansion includes only even powers of the perturbation. On the other hand, $e_{ab}^{\textrm{BGE2},2nd}$ can be interpreted
as 32 quadruple-excitation Goldstone diagrams which, however, are both disconnected. The $e_{ab}$-coupling effect in BGE2 does
therefore not produce higher-order connected Goldstone diagrams. 
Using the so-called Hugenholtz diagram rule\cite{brian:1975A}, these 32 quadruple-excitations can be represented by two Hugenholtz diagrams, which are shown in 
fig.~\ref{Fig:disconnected-4th-diagram}.

Recently, a self-consistent Green's function schem was proposed at 2nd arder as well\cite{phillips:2014A}.
This self-consistent second-order self-energy method also exhibits promising performance for systems with strong correlation. 
It would be very interesting to compare the diagrams of the Green's function theory with BGE2 in the future. 

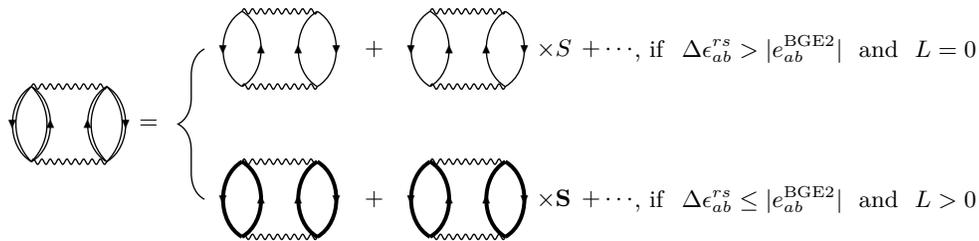
\begin{figure*}[t]
    \pgfdeclaredecoration{complete sines}{initial}
    {
        \state{initial}[
            width=+0pt,
            next state=sine,
            persistent precomputation={\pgfmathsetmacro\matchinglength{
                \pgfdecoratedinputsegmentlength / int(\pgfdecoratedinputsegmentlength/\pgfdecorationsegmentlength)}
                \setlength{\pgfdecorationsegmentlength}{\matchinglength pt}
            }] {}
        \state{sine}[width=\pgfdecorationsegmentlength]{
            \pgfpathsine{\pgfpoint{0.25\pgfdecorationsegmentlength}{0.5\pgfdecorationsegmentamplitude}}
            \pgfpathcosine{\pgfpoint{0.25\pgfdecorationsegmentlength}{-0.5\pgfdecorationsegmentamplitude}}
            \pgfpathsine{\pgfpoint{0.25\pgfdecorationsegmentlength}{-0.5\pgfdecorationsegmentamplitude}}
            \pgfpathcosine{\pgfpoint{0.25\pgfdecorationsegmentlength}{0.5\pgfdecorationsegmentamplitude}}
    }
        \state{final}{}
    }
    \tikzset{wave/.style={decorate, decoration={complete sines, amplitude=2pt, segment length=3pt}}}
	\begin{tikzpicture}[scale=\myscale]
		\path (-0.2,0.5) coordinate (b1);
		\path (2.00,0.5) coordinate (b2);
		\path (2.50,1.5) coordinate (b3);
		\path (2.5,-0.5) coordinate (b4);
		
		\path (0.2,0.0) coordinate (b2-a)
		      (0.2,1.0) coordinate (b3-a);
		\path (1.2,0.0) coordinate (b2-b)
		      (1.2,1.0) coordinate (b3-b);
		\path (5.00,1.5) coordinate (b5);
		\path (6.80,1.5) coordinate (b6);
		\path (7.4,1.5) coordinate (b7);
		\path (8.1,1.5) coordinate (b8);
		\path (3.0,1.0) coordinate (b4-a)
		      (3.0,2.0) coordinate (b5-a);
		\path (4.0,1.0) coordinate (b4-b)
		      (4.0,2.0) coordinate (b5-b);
		\path (5.5,1.0) coordinate (b6-a)
		      (5.5,2.0) coordinate (b7-a);
		\path (6.5,1.0) coordinate (b6-b)
		      (6.5,2.0) coordinate (b7-b);
		\path (5.00,-0.5) coordinate (a5);
		\path (6.80,-0.5) coordinate (a6);
		\path (7.40,-0.5) coordinate (a7);
		\path (8.10,-0.5) coordinate (a8);
		\path (3.0,-1.0) coordinate (a4-a)
		      (3.0,0.0) coordinate (a5-a);
		\path (4.0,-1.0) coordinate (a4-b)
		      (4.0,0.0) coordinate (a5-b);
		\path (5.5,-1.0) coordinate (a6-a)
		      (5.5,0.0) coordinate (a7-a);
		\path (6.5,-1.0) coordinate (a6-b)
		      (6.5,0.0) coordinate (a7-b);
		\draw[line width=0.5] (b2-a) edge[bend left=60] node {\fontsize{\pta}{1em}\selectfont $\blacktriangledown$} (b3-a);
		\draw[line width=0.5] (b2-a) edge[bend left=45] (b3-a);
		\draw[line width=0.5] (b2-a) edge[bend right=60] node {\fontsize{\pta}{1em}\selectfont $\blacktriangle$} (b3-a);
		\draw[line width=0.5] (b2-a) edge[bend right=45] (b3-a);
		\draw[line width=0.5, wave] (b2-a) -- (b2-b);
		\draw[line width=0.5, wave] (b3-a) -- (b3-b);
		\draw[line width=0.5] (b2-b) edge[bend left=60] node {\fontsize{\pta}{1em}\selectfont $\blacktriangle$} (b3-b);
		\draw[line width=0.5] (b2-b) edge[bend left=45] (b3-b);
		\draw[line width=0.5] (b2-b) edge[bend right=60] node {\fontsize{\pta}{1em}\selectfont $\blacktriangledown$} (b3-b);
		\draw[line width=0.5] (b2-b) edge[bend right=45] (b3-b);
		\node[left] at (b2) {\fontsize{\ptb}{1em}\selectfont $ = $};
		\draw[line width=0.5, decorate, decoration={brace, amplitude=10pt, mirror},xshift=0pt,yshift=0pt] (b3) -- (b4);
		\draw[line width=0.5] (b4-a) edge[bend left=60] node {\fontsize{\pta}{1em}\selectfont $\blacktriangledown$} (b5-a);
		\draw[line width=0.5] (b4-a) edge[bend right=60] node {\fontsize{\pta}{1em}\selectfont $\blacktriangle$} (b5-a);
		\draw[line width=0.5, wave] (b4-a) -- (b4-b);
		\draw[line width=0.5, wave] (b5-a) -- (b5-b);
		\draw[line width=0.5] (b4-b) edge[bend left=60] node {\fontsize{\pta}{1em}\selectfont $\blacktriangle$} (b5-b);
		\draw[line width=0.5] (b4-b) edge[bend right=60] node {\fontsize{\pta}{1em}\selectfont $\blacktriangledown$} (b5-b);
		\node[left] at (b5) {\fontsize{\ptb}{1em}\selectfont $ + $};
		\draw[line width=0.5] (b6-a) edge[bend left=60] node {\fontsize{\pta}{1em}\selectfont $\blacktriangledown$} (b7-a);
		\draw[line width=0.5] (b6-a) edge[bend right=60] node {\fontsize{\pta}{1em}\selectfont $\blacktriangle$} (b7-a);
		\draw[line width=0.5, wave] (b6-a) -- (b6-b);
		\draw[line width=0.5, wave] (b7-a) -- (b7-b);
		\draw[line width=0.5] (b6-b) edge[bend left=60] node {\fontsize{\pta}{1em}\selectfont $\blacktriangle$} (b7-b);
		\draw[line width=0.5] (b6-b) edge[bend right=60] node {\fontsize{\pta}{1em}\selectfont $\blacktriangledown$} (b7-b);
		\node[right] at (b6) {\fontsize{\ptb}{1em}\selectfont $ \times S$};
		\node[right] at (b7) {\fontsize{\ptb}{1em}\selectfont $ + \cdots$};
		\node[right] at (b8) {\fontsize{\ptb}{1em}\selectfont \textrm{, if } 
		$\Delta \epsilon_{ab}^{rs}>|e_{ab}^{\textrm{BGE2}}|$ \textrm{ and } $L=0$};
		\draw[line width=1.5] (a4-a) edge[bend left=60] node {\fontsize{\pta}{3em}\selectfont $\blacktriangledown$} (a5-a);
		\draw[line width=1.5] (a4-a) edge[bend right=60] node {\fontsize{\pta}{3em}\selectfont $\blacktriangle$} (a5-a);
		\draw[line width=0.5, wave] (a4-a) -- (a4-b);
		\draw[line width=0.5, wave] (a5-a) -- (a5-b);
		\draw[line width=1.5] (a4-b) edge[bend left=60] node {\fontsize{\pta}{1em}\selectfont $\blacktriangle$} (a5-b);
		\draw[line width=1.5] (a4-b) edge[bend right=60] node {\fontsize{\pta}{1em}\selectfont $\blacktriangledown$} (a5-b);
		\node[left] at (a5) {\fontsize{\ptb}{1em}\selectfont $ + $};
		\draw[line width=1.5] (a6-a) edge[bend left=60] node {\fontsize{\pta}{1em}\selectfont $\blacktriangledown$} (a7-a);
		\draw[line width=1.5] (a6-a) edge[bend right=60] node {\fontsize{\pta}{1em}\selectfont $\blacktriangle$} (a7-a);
		\draw[line width=0.5, wave] (a6-a) -- (a6-b);
		\draw[line width=0.5, wave] (a7-a) -- (a7-b);
		\draw[line width=1.5] (a6-b) edge[bend left=60] node {\fontsize{\pta}{1em}\selectfont $\blacktriangle$} (a7-b);
		\draw[line width=1.5] (a6-b) edge[bend right=60] node {\fontsize{\pta}{1em}\selectfont $\blacktriangledown$} (a7-b);
		\node[right] at (a6) {\fontsize{\ptb}{1em}\selectfont $ \times \mathbf{S}$};
		\node[right] at (a7) {\fontsize{\ptb}{1em}\selectfont $ + \cdots$};
		\node[right] at (a8) {\fontsize{\ptb}{1em}\selectfont \textrm{, if } 
		$\Delta \epsilon_{ab}^{rs}\le |e_{ab}^{\textrm{BGE2}}|$ \textrm{ and } $L>0$};

	\end{tikzpicture}
	\caption{\label{Fig:diagram-BGE2-diagonal}
		The perturbative expansion of the direct term in the BGE2 correlation. Double lines in the BGE2 diagram represent a correction to the double 
		excitation energies due to the $e_{ab}$-coupling effect, which should be solved iteratively. Thin lines for the case of $L=0$ are the particles and holes
		for the standard many-body perturbation theory (top). However, thick lines in the perturbation expansion for the $L>0$ case suggest a constant level-shift
		$L$ to the double excitation energies (bottom).
		The squiggly lines refer to the bare Coulomb interaction.
	}
\end{figure*}

Now we consider the case of $\Delta\epsilon_{ab}^{rs}\le |e_{ab}^{\textrm{BGE2}}(\lambda)|$ where the (near)-degeneracy effects are dominant. 
As mentioned above, to guarantee a convergent geometric expansion, a positive level-shift $L>-\frac{1}{2}(\Delta\epsilon_{ab}^{rs}+e_{ab}^{\textrm{BGE2}}(\lambda))$ 
is required. By inserting the binomial series $\left( e_{ab}^{\textrm{BGE2}}+L \right)^n$ into eq.~\ref{Eq:BGE2am} we obtain the corresponding geometric series
\begin{widetext}
\begin{equation}
	\label{Eq:BGE2ame2}
	e_{ab}^{\textrm{BGE2}}(\lambda)=-\sum_{n}^{\infty}\sum_{m}^{n}{n \choose m}L^{n-m}
		\sum_{r<s}^{unocc}\frac{\lambda^2\left|\left<\phi_{a}\phi_{b}||\phi_{r}\phi_{s}\right>\right|^2}
		{(\Delta\epsilon_{ab}^{rs}+L)^{n+1}}e_{ab}^{\textrm{BGE2}}(\lambda)^m .
\end{equation}
\end{widetext}
We note that the $L$=0 case in eq.~\ref{Eq:BGE2ame} is a special case of eq.~\ref{Eq:BGE2ame2}. For positive level shifts, it can be easily proven that 
the first two expansion terms of eq.~\ref{Eq:BGE2ame2} are the same as in eqs.~\ref{Eq:BGE2-1st} and \ref{Eq:BGE2-2nd}, only that the level-shift $L$ 
appears in the denominator. 
We plot the perturbative expansion of the BGE2 correlation in fig.~\ref{Fig:diagram-BGE2-diagonal}. For the $L$=0 case, the BGE2 correlation can be
interpreted based on the standard perturbative expansion. However, if $\Delta\epsilon_{ab}^{rs}\le |e_{ab}^{\textrm{BGE2}}(\lambda)|$, 
a positive level shift is required to guarantee a well-defined perturbative expansion of the BGE2 correlation. In fig.~\ref{Fig:diagram-BGE2-diagonal},
we show the first- and second-order geometric expansion of the direct term of the BGE2 correlation. The exchange part can be expanded in the same way.
Thick lines are introduced to represent a positive level-shift $L$ to the double excitation energies $\Delta\epsilon_{ab}^{rs}$. For both cases, higher
order excitations are not involved, but only rescale weights of the existing contributions. The $\mathbf{S}$ is the modified $S$ with a positive 
level-shift $L$:
\begin{equation}
	\begin{split}
		\mathbf{S}=&\sum_{r<s}^{unocc}\frac{\left|\left<\phi_{a}\phi_{b}||\phi_{r}\phi_{s}\right>\right|^2}
		{\left(\Delta\epsilon_{ab}^{rs}+L\right)^2}
	\end{split}
\end{equation}

The divergence of M\o{}ller-Plesset (MP) and G\"orling-Levy (GL) perturbation theories has been widely discussed in quantum 
chemistry\cite{jorgensen:1996A,sherrill:2000A,jorgensen:1996B,jorgensen:2000A,herman:2009A}. Small-gap systems with strong 
(near)-degeneracy effects are, of course, one kind of failure, as the perturbation energy diverges at any finite order. 
For non-degenerate systems where $\Delta\epsilon_{ab}^{rs}$ is not exactly zero, the MP (or GL) perturbation expansion does not diverge.
However, Leininger {\it et al.} found that 
the perturbation expansion also does not converge toward the exact solution even for very simple systems such as Ne, F$^-$, and 
Cl$^-$\cite{sherrill:2000A}. The individual terms in the perturbative expansion of these systems do not diverge, but exhibit an oscillatory 
divergence with increasing order\cite{jorgensen:1996A,sherrill:2000A}.
From a mathematical point of view the MP expansion fails, if the single
reference, e.g., HF or KS, is far from the exact ground state \cite{jorgensen:1996B,jorgensen:2000A,herman:2009A}. However, there is no
simple diagnostic tool to determine when a multi-reference problem breaks the MP expansion.

In this work we propose to use the condition $\Delta\epsilon_{ab}^{rs}\le |e_{ab}^{\textrm{BGE2}}(\lambda)|$ as a simple criterion to
judge the divergence of a single-reference perturbation method. If the HOMO-LUMO gap of a given single reference is smaller than the 
absolute value of the corresponding BGE2 electron-pair correlation energy, it is not advisable to use a perturbative method based on 
this single reference, because the BGE2 correlation cannot be expanded directly without a proper level-shift $L$ (eq.~\ref{Eq:BGE2am}). 
The value of the level shift $L$ then quantifies the multi-reference nature of each electron pair in the investigated systems.

\subsection{\label{SSec:Num} H$_2$ and H$_2^{+}$ dissociation}

Our BGE2 $xc$ functional encompasses the exact exchange and the BGE2 correlation term (eqs.~\ref{Eq:BGE2a} and \ref{Eq:BGE2total})
\begin{equation}
	E_{xc}^{\textrm{BGE2}}=E_{x}^{\textrm{EX}}+E_{c}^{\textrm{BGE2}}(\lambda=1).
\end{equation}
It has been implemented in the Fritz Haber Institute \emph{ab initio molecular simulations} (FHI-aims) code
package\cite{blum:2009A,igor:2013A}. Due to its simple sum-over-state formula, BGE2 has the same computational scaling as standard PT2 in terms of both
time and memory. Although the $e_{ab}$-coupling requires an iterative solution,  convergence is fast in our experience, and an accuracy of $10^{-8}$ 
Hartree can be achieved within a few iterations.

\begin{figure}[!tbp]
    \includegraphics[scale=0.425]{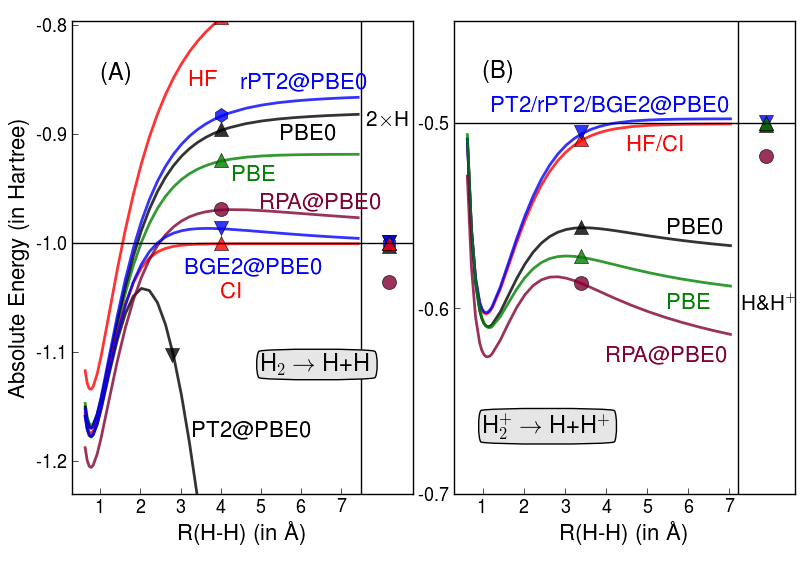}
    \caption{\label{Fig:PES} 
	H$_2$ (A) and H$_2^+$ (B)  dissociation curves without breaking spin symmetry. 
	All calculations, including the configuration interaction method with singles and doubles 
	(CISD), have been carried out with FHI-aims\cite{blum:2009A} and the NAO-VCC-5Z basis set\cite{igor:2013A}. For one and two electron systems,
	CISD provides the exact curves, which are thus denoted as CI. The HF, PBE and PBE0 results are
	obtained self-consistently. The HF orbitals are employed to evaluate the CISD results, and the PT2, RPA, rPT2, and BGE2 calculations are on top of
	a PBE0 reference.
	In the smaller panel for H$_2$ dissociation, the total energies of two isolated Hydrogen atoms are plotted for each method.
	And for H$_2^+$ dissociation, the smaller panel shows the total energies of one isolated Hydrogen atom.
	}
\end{figure}

\begin{figure}[!tbp]
    \includegraphics[scale=0.425]{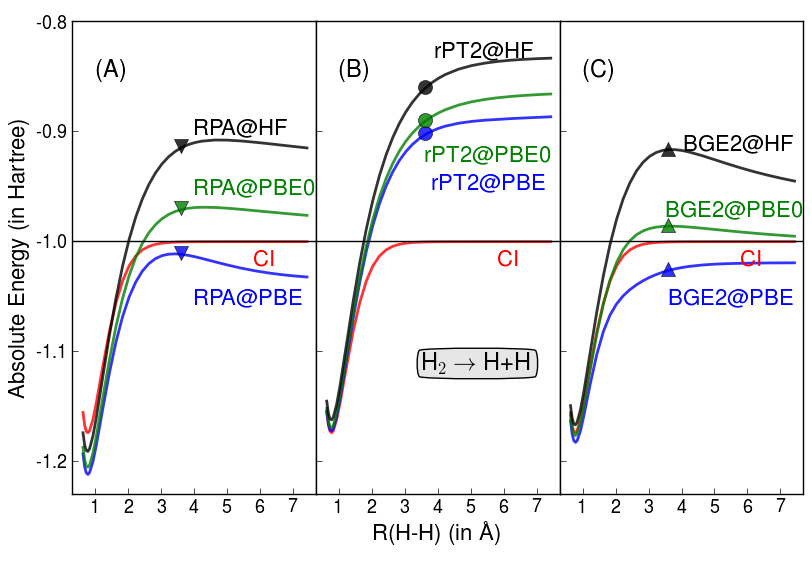}
    \caption{\label{Fig:REF} 
	The H$_2$ dissociations calculated by RPA (A), rPT2 (B) and BGE2 (C). The nomenclature adopted here: F@SC is the advanced functionals (F), 
	i.e.\ RPA, rPT2, and BGE2, respectively, evaluated with the orbitals of different self-consistent (SC) schemes, i.e.\ HF, PBE0, and PBE. }
\end{figure}

In fig.~\ref{Fig:PES} we plot the H$_2$ and H$_2^+$ dissociation curves for various methods (BGE2, PBE, PBE0, PT2, RPA, and rPT2).  
All results are obtained with input KS orbitals from a PBE0 calculation \cite{perdew:1996B,adamo:1999A,scuseria:1999A}. 
All calculations, including the CISD reference, are carried out with FHI-aims using the NAO-VCC-5Z basis set\cite{igor:2013A,ihrig:2015A}.
In fig.~\ref{Fig:REF} we show the same curves for RPA, rPT2 and BGE2 for different starting points (PBE, PBE0 and HF). 

In the following we will analyse the performance of the different approaches shown in fig.~\ref{Fig:PES} class by class. In
the dissociation of H$_2^+$ only non-local exact exchange is required and correlation is absent, while dissociated H$_2$ contains 
strong (near)-degeneracy static correlation, which current DFT methods typically underestimate. As illustrated in fig.~\ref{Fig:PES}, 
the PBE0 functional fails in both cases, yielding a heavy one-electron ``self-correlation'' error (around 66 mHartree in the H$_2^+$ 
dissociation limit) and a significant underestimation of the
(near)-degeneracy static correlation limit (around 119 mHartree in the H$_2$ dissociation limit).

Fifth-level correlation functionals, e.g., PT2, RPA and rPT2, are fully non-local. 
Fig.~\ref{Fig:PES} reveals that PT2 is exact for one-electron systems such as H$_2^+$ dissociation, in agreement with previous investigations\cite{yang:2011A,
gorling:2011A,furche:2013A,scuseria:2010A,scuseria:2010B}, but completely fails for heavily stretched H$_2$, where  static correlation becomes dominant. 
RPA exhibits the opposite behavior. It is accurate in the H$_2$ dissociation limit, but its severe one-electron ``self-interaction'' or delocalization 
error\cite{Mori-Sanchez/etal:2012,Hellgren/etal:2015} affects H$_2^+$ dissociation adversely.  Adding the SOSEX term to RPA removes the one-electron 
self-interaction error again such that rPT2 dissociates H$_2^+$ correctly, but simultaneously the performance for H$_2$ deteriorates \cite{rinke:2013A}.
Henderson {\it et al.} have ascribed this behavior to a removal of static correlation by the SOSEX term \cite{scuseria:2010A}. 

Fig.~\ref{Fig:PES} reveals that the BGE2 functional is free of one-electron ``self-correlation'' and thus dissociates H$_2^+$ correctly. It also delivers 
the correct H$_2$ dissociation limit and reduces the incorrect repulsive ``bump'' that RPA exhibits at intermediate bond lengths. We attribute this 
consistent improvement to three factors. 1) The one-electron ``self-correlation'' error is removed at the PT2 level. 2) (Near)-degeneracy static
correlation (for two electrons) is incorporated by the $e_{ab}$-coupling mechanism at the second-order perturbation level without having to invoke 
higher-order Goldstone diagrams.
3) Due to the systematic nature of the approximations we made, we can trace the repulsive ``bump'' back to the 2nd order approximation, when we went 
from the full BGE to BGE2. In other words higher-order pp-ladder diagrams will alleviate the ``bump'' \cite{olsen:2014A}. 

In fig.~\ref{Fig:REF} we illustrate the starting-point dependence of RPA, rPT2 and BGE2 by evaluating all three approaches for a PBE, a PBE0 and a HF
reference. The starting-point dependence in RPA is quite pronounced, which is akin to the much investigated starting-point dependence in the corresponding 
$GW$ approach for excitation spectra \cite{Rinke/etal:2005,Fuchs/etal:2007,Marom/etal:2012,Bruneval/Marques:2013}. 
Judging by  fig.~\ref{Fig:REF}, the starting-point dependence of rPT2 and BGE2 appears to be as pronounced as for RPA. Although it remains to be 
seen in the future, if this statement can be generalized. 

\section{\label{Sec:Properties} Promising properties of the BGE2 correlation functional}
In wave-function theory, there is a systematic way to improve the theoretical approach for the electronic correlation energy \cite{szabo:1996A,nesbet:1971A,
pople:1976A,pople:1987A}. However, it is challenging to design methods that fulfil a certain number of exact conditions and constraints \cite{pople:1987A}.
One requirement is that the solution for one- and two-electron systems such as H$_2$/H$_2^+$ is exact. Another requirement is size consistency \cite{szabo:1996A,
bartlett:1981A,hirata:2011A},  i.e., the ground-state total energy itself is an extensive quality, which should be asymptotically proportional to the system size 
\cite{szabo:1996A}. These two criteria are widely used to judge the universal applicability of a given theoretical method from small isolated molecules to  
extensive solids. We will analyse analytically, how well BGE2 fares for H$_2$ in a minimal basis.

\subsection{\label{SSec:H2min} H$_2$ in minimal basis}
The BGE2 approximation is not exact for two-electron systems. However, in Sec.~\ref{SSec:Num}, we reported a significant improvement of BGE2 over PT2 and RPA for 
H$_2$ dissociation. In this section, we analyze the BGE2 correlation functional (eq.~\ref{Eq:BGE2a}) and its correlation potential (eq.~\ref{Eq:BGE2Vab}) for 
H$_2$  in a minimal basis. We demonstrate that the BGE2 approximation captures essential features of the adiabatic connection path that current state-of-the-art
approximations do not.

\begin{figure}[!tbp]
    \includegraphics[scale=0.425]{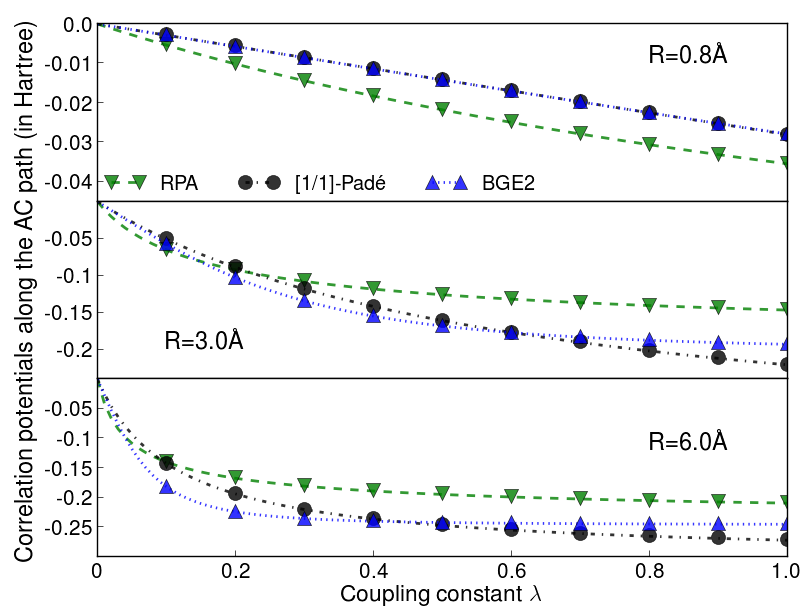}
    \caption{\label{Fig:CP} 
	Correlation potentials of H$_2$ in a minimal basis at the equilibrium geometry (R=0.8\AA, upper panel), an intermediate bond distance
	(R=3.0\AA, middel panel), and a large bond distance (R=6.0\AA, lower panel) for RPA, the [1/1]-Pad\'e model, and BGE2.
	The parameters in the [1/1]-Pad\'e model 
	are determined by fixing the initial slope to 2$E_c^{\textrm{PT2}}$
	and the correlation energy to $E_{c}^{\textrm{BGE2}}$. 
	}
\end{figure}

The minimal basis of H$_2$ consists of one bonding ($\psi_+$) and one anti-bonding ($\psi_-$) KS orbital determined 
by the $D_{\infty h}$ symmetry:
\begin{equation}
	\psi_{\pm}(\boldsymbol{r})=\frac{\phi_{1s}(\boldsymbol{r}-R_1)\pm \phi_{1s}(\boldsymbol{r}-R_2)}{\sqrt{2+2S_{R1,R2}}},
\end{equation}
where $R_1$ and $R_2$ are the two atomic positions, $\phi_{1s}(\boldsymbol{r}-R_{1/2})$ is the normalized $1s$ atomic orbital located at the each hydrogen atom, 
and $S_{R_1,R_2}$ is the overlap integral of the two atomic orbitals. In this minimal basis representation, the BGE2 correlation energy (eq.~\ref{Eq:BGE2a})
and its potential with respect to the coupling constant $\lambda$ (eq.~\ref{Eq:BGE2Vab}) can be derived analytically
\begin{equation}
	\label{Eq:BGE2_H2_minimal}
	\begin{split}
		E_{c}^{\textrm{BGE2}}(\lambda)=&\frac{B-\sqrt{B^2+4\lambda^2A}}{2}\\
    	V_{c}^{\textrm{BGE2}}(\lambda)=&-\frac{2\lambda A}{\sqrt{B^2+4\lambda^2A}}
	\end{split}
\end{equation}
where $A=|\left<\psi_{+}\psi_{+}|\psi_{-}\psi_{-}\right>|^2$ and $B=\Delta \epsilon_{++}^{--}$. Now we see that several important features of the AC path are
captured by the BGE2 approximation:

i) When $\lambda\rightarrow 0$, we have
\begin{equation}
	\label{Eq:initial-slope}
	V_{c}^{\textrm{BGE2}'}(0)=\left.\frac{V_{c}^{\textrm{BGE2}}(\lambda)}{\partial \lambda}\right|_{\lambda=0}=-2\frac{A}{B}=2E_{c}^{\textrm{PT2}}
\end{equation}
which shows that the initial slope of the BGE2 correlation potential is twice the energy in PT2. Comparing to the exact condition,
i.e. $V_{c}^{'}(0)=2E_{c}^{\textrm{GL2}}$, the single-excitation contribution is ignored during the first approximation in this work (eqs.~\ref{Eq:GF0_1} and \ref{Eq:GF0}). 

ii) In the H$_2$ dissociation limit ($|R_1-R_2|\rightarrow \infty$ and $B=\Delta \epsilon_{++}^{--}\rightarrow 0$) we find for 
the initial slope $V_{c}^{\textrm{BGE2}'}=-2\frac{A}{B}\rightarrow \infty$. The BGE2 correlation potential itself tends to a constant value that is 
independent of $\lambda$ thanks to the $e_{ab}$-coupling effect:
\begin{equation}
	\label{Eq:H2min-exact-correlation}
	\left.V_{c}^{\textrm{BGE2}}(\lambda)\right|_{|R_1-R_2|\rightarrow\infty}\rightarrow
	-\left<\phi_{1s}\phi_{1s}|\phi_{1s}\phi_{1s}\right>.
\end{equation}
The coupling-constant integration (eq.~\ref{Eq:AC}) is trivial to carry out and the correlation energy $E_{c}^{\textrm{BGE2}}$ has 
the same value as $V_{c}^{\textrm{BGE2}}$. This is the exact correlation energy in the minimal basis \cite{szabo:1996A}. 
In conjunction with the exact exchange energy, it completely cancels out the undesired error originating from the Hartree 
approximation, and thus guarantees the correct dissociation limit.

If we do not make the fourth approximation, i.e. to remove the first-order perturbation term $e_{ab}^{1st}$ in the denominator of eq.~\ref{Eq:BGE2}, the parameter $B$ now equals $\Delta\epsilon_{++}^{--}+e_{++}^{1st}(\lambda)$. By using the definitions of
$\hat{v}_{\lambda}$ and $e_{ab}^{1st}$ (eqs.~\ref{Eq:CS} and \ref{Eq:BGE-1st}) we have $e_{++}^{1st}(\lambda)$ in the H$_2$ dissociation limit
\begin{equation}
	\label{Eq:H2min-BGE2-1st}
	e_{++}^{1st}(\lambda)\left|_{|R_1-R_2|\rightarrow\infty}\right.\approx 
	-\lambda\left<\phi_{1s}\phi_{1s}|\phi_{1s}\phi_{1s}\right>
\end{equation}
Here, for simplicity, we neglect the exchange $v_{x}(\boldsymbol{r}_i)$ and the scaled correlation potential $v_{c}(\boldsymbol{r}_i,\alpha)$ (eq.~\ref{Eq:CS}), since they are small compared to the Hartree energy $E_{H}$. The resulting correlation energy $E_{c}=-\frac{2}{\sqrt{5}}\left<\phi_{1s}\phi_{1s}|\phi_{1s}\phi_{1s}\right>$ recovers only 90\% of the exact value in the minimal basis (eq.~\ref{Eq:H2min-exact-correlation}). This motivates {\it a posteriori} why we made the second approximation, i.e. omitted the single excitation contribution. In practice this single excitation contribution is non-zero in a DFT framework. However, our minimal basis consideration shows that if we were to include it, H$_2$ would no longer dissociate correctly, unless we would also include higher-order particle-particle ladder diagrams (the third approximation), which would significantly increase the computational cost. The BGE2 correlation functional proposed in this work is thus the simplest approximation that 
provides the exact H$_2$ dissociation in the minimal basis. Any further improvement over BGE2 has to simultaneously deal with all approximations in a proper and balanced way.

iii) The second-order derivative of the BGE2 correlation energy $V_{c}''=\partial V_{c}(\lambda)/\partial \lambda$ is
\begin{equation}
	V_{c}^{\textrm{BGE2}''}(\lambda)=\frac{24A^2B^{-3}\lambda}{\left(1+4\lambda^2A/B^2\right)^{5/2}}.
\end{equation}
As $A$ and $B$ are positive, $V_{c}^{\textrm{BGE2}''}(\lambda)\ge 0$,
indicating that the first  derivative of  $V_{c}^{\textrm{BGE2}}$ is monotonically increasing within $0\le \lambda\le 1$,
\begin{equation}
	V_{c}^{\textrm{BGE2}'}(\lambda+|\delta \lambda|)\ge V_{c}^{\textrm{BGE2}'}(\lambda).
\end{equation}
Considering that $V_{c}^{\textrm{BGE2}'}(\lambda) \le 0$ for both $\lambda=0$ and $\lambda=1$, $V_{c}^{\textrm{BGE2}}(\lambda)$
captures the convex shape of the exact AC correlation path \cite{ernzerhof:1996A,yang:2007A,helgaker:2010A}.

v) The $\lambda$-dependent AC path is closely connected to the behavior under uniform density scaling \cite{perdew:1993A}, for which  the low-density
limit is related to the strong correlation limit $\lambda\rightarrow\infty$. The exact correlation functional should reach a finite value in the strong 
correlation limit \cite{perdew:1993A,ernzerhof:1996A,helgaker:2010A}, which is satisfied by the BGE2 model according to eq.~\ref{Eq:BGE2_H2_minimal} 
($V_{c}\rightarrow -
 \sqrt{A}$ when $\lambda\rightarrow\infty$). 

Figure \ref{Fig:CP} presents the correlation potentials of RPA and BGE2 for H$_2$ dissociation in the minimal basis. 
At the equilibrium geometry (R=0.8\AA) where the energy gap is large, the BGE2 model 
exhibits a quasi-linear behavior. For stretched geometries (R= 3.0 and 6.0 \AA), the $e_{ab}$-coupling then bends the correlation
potential, whereas RPA overestimates the initial slopes. The BGE2 correlation potentials are similar to those of configuration interaction calculations with
a quadrupole-$\zeta$
basis set \cite{helgaker:2010A}. 

We also include the common [1/1]-Pad\'e AC model\cite{ernzerhof:1996A,yang:2007A} 
\begin{equation}
	\label{Eq:Pade}
	V_{c}^{\textrm{Pad\'e}}(\lambda)=\frac{a\lambda}{1+b\lambda},\ E_{c}^{\textrm{Pad\'e}}=\frac{a}{b}-\frac{a\ln(1+b)}{b^2}.
\end{equation}
The two parameters are determined by fixing the initial slope to 2$E_{c}^{\textrm{PT2}}$ and the correlation energy to $E_{c}^{\textrm{BGE2}}$.
As illustrated in fig.\ \ref{Fig:CP}, the [1/1]-Pad\'e formula describes the AC path at the equilibrium geometry well, but exhibits a 
tendency to underestimate the curvature and overestimate the correlation potential at stretched geometries. A better agreement
can be expected by using a more sophisticated [2/2]-Pad\'e formula \cite{ernzerhof:1996A,yang:2007A}.

\subsection{\label{SSec:SC} Size consistency for the ground-state energy calculation}
Recently, the application of advanced correlation methods, e.g., MP2 \cite{kresse:2009A,joost:2012A,joost:2013A}, RPA \cite{joost:2013A,kresse:2010A,RPAreview,
kresse:2014A}, and coupled-cluster theories \cite{andreas:2013A,scuseria:2014A,scuseria:2014B}, in materials science has attracted increased attention.  In this paper,
we adopt the K-dependence  criterion \cite{hirata:2011A,hirata:2010A,march:1967A,calais:1995A,bartlett:1997A} to demonstrate the size consistency of the BGE2 
correlation functional and its applicability to complex extended materials. The advantage and usage of the  K-dependence criterion has been discussed comprehensively 
in Ref.~\onlinecite{hirata:2011A}. In short, the number ($K$) of k-points in the Brillouin zone is a direct measure of system size in periodic boundary conditions. 
Then a size-consistent method must have an asymptotic $K^1$ dependence. 

Before we turn to BGE2, we first discuss Brillouin-Wigner second-order perturbation theory (BW2) \cite{march:1967A,bartlett:2009A}. The size consistency
of BW2 has been disproved in Ref.~\onlinecite{hirata:2011A}. This helps us to better understand the size consistency  of BGE2 which shares a very similar 
sum-over-state formula as BW2.

In periodic boundary conditions, the BW2 correlation is given by
\begin{equation}
	\label{Eq:BW2}
	\begin{split}
	    E_{c}^{\textrm{BW2}}
		=&\sum_{a<b}\sum_{r<s}\sum_{k_bk_rk_s}
	    \frac{\left|\left<\phi_{ak_a}\phi_{bk_b}||\phi_{rk_r}\phi_{sk_s}\right>\right|^2}
	    {E_{c}^{\textrm{BW2}}+E_{x}^{\textrm{EX}}-\Delta\epsilon_{ak_abk_b}^{rk_rsk_s}}\\
		\approx&\sum_{a<b}\sum_{r<s}\sum_{k_bk_rk_s}
	    \frac{\left|\left<\phi_{ak_a}\phi_{bk_b}||\phi_{rk_r}\phi_{sk_s}\right>\right|^2}
	    {E_{c}^{\textrm{BW2}}-\Delta\epsilon_{ak_abk_b}^{rk_rsk_s}}\\
	\end{split}
\end{equation}
Here, $\phi_{ik_i}$ is a canonical HF or KS spin-orbital in the $i$th band with wave vector $k_i$. Due to momentum conservation, the summation only goes over
three wave vectors ($k_b,k_r,k_s$), giving rise to a factor $K^3$. It can be proven that $\left|\left<\phi_{ak_a}\phi_{bk_b}||\phi_{rk_r}\phi_{sk_s}\right>\right|^2$
exhibits an asymptotic $K^{-2}$ dependence. $E_{x}^{\textrm{EX}}$ and $\Delta\epsilon_{ak_abk_b}^{rk_rsk_s}$ scale as $K^1$ and $K^0$, respectively \cite{hirata:2011A}.
If the denominator scales the same as $E_{x}^{\textrm{EX}}$ (the first line in eq.~\ref{Eq:BW2}), the overall scaling of $E_{c}^{\textrm{BW2}}$ becomes $K^0$, which
would not be size consistent. Removing $E_{x}^{\textrm{EX}}$ from the denominator (second line in eq.~\ref{Eq:BW2}) changes the scaling of $E_{c}^{\textrm{BW2}}$ 
to $K^{1/2}$, which is still not size consistent. These non-physical size dependences make the second-order energy per unit cell, $E_{c}^{\textrm{BW2}}/K$, go to 
zero as $K\rightarrow\infty$. It has been argued that the presence of an extensive quantity, $E_{c}^{\textrm{BW2}}+E_{x}^{\textrm{EX}}$, in the denominator is 
responsible for the lack of size consistency\cite{hirata:2011A,bartlett:2009A}.

The BGE2 correlation energy (eqs.~\ref{Eq:BGE2a} and \ref{Eq:BGE2total}) in periodic boundary conditions takes the form
\begin{equation}
	\label{Eq:BGE2periodic}
	\begin{split}
	    e_{ab}^{\textrm{BGE2}}(\lambda)
		=&\sum_{r<s}\sum_{k_rk_s}
	    \frac{\lambda^2\left|\left<\phi_{ak_a}\phi_{bk_b}||\phi_{rk_r}\phi_{sk_s}\right>\right|^2}
	    {e_{ab}^{\textrm{BGE2}}(\lambda)-\Delta\epsilon_{ak_abk_b}^{rk_rsk_s}}\\
		E_{c}^{\textrm{BGE2}}(\lambda)=&\sum_{a<b}\sum_{k_b}e_{ab}^{\textrm{BGE2}}(\lambda) .
	\end{split}
\end{equation}
Compared to the BW2 correlation energy, the only difference in BGE2 is the appearance of the correlation coupling for each electron pair $ab$, 
i.e.\ the $e_{ab}$-coupling. 
Following a similar approach as for BW2, it is easy to prove that the electron-pair correlation term $e_{ab}^{\textrm{BGE2}}$ scales as $K^0$~\cite{hirata:2011A}. 
Due to momentum conservation the summation in  the BGE2 correlation energy $E_{c}^{\textrm{BGE2}}$ only runs over one wave vectors $k_b$ (see eq.~\ref{Eq:BGE2periodic}). Therefore, $E_{c}^{\textrm{BGE2}}$ scales as $K^1$ and thus is size consistent.

\subsection{\label{SSec:OIV} Orbital invariance}

It should be stated that an electron-pair approximation such as BGE2 does not have a derived wave function. It thus breaks another important feature:
the orbital invariance \cite{szabo:1996A}, i.e.\ the total energies cannot be determined uniquely with respect to rotations among occupied and/or 
unoccupied orbitals. However, the simple sum-over-state formula of BGE2 (eqs. \ref{Eq:BGE2a} and \ref{Eq:BGE2periodic}) clearly suggests that the 
$e_{ab}$-coupling decays very quickly to standard PT2 when the energy difference becomes larger. Therefore, we expect that this orbital invariance 
deficiency does not affect real applications. We leave a detailed examination of this issue to the future work.

\section{Conclusion}
In this work, we present a new insight into the H$_2$/H$_2^+$ challenge in DFT. We establish the Bethe-Goldstone equation 
in the context of DFT through the adiabatic-connection approach. BGE is the simplest approximation to provide the exact solution for one- and two-electron 
systems. We propose a simple orbital-dependent correlation functional, BGE2, by terminating the BGE expansion at the second order, but reversing the 
$e_{ab}$-coupling effect in BGE. BGE2 has a similar sum-over-state formula as the standard PT2, thus sharing the same computational scaling as PT2
in terms of both time and memory.
We demonstrates that the $e_{ab}$-coupling iteration procedure at the second-order expansion does not invoke higher-order connected Goldstone 
diagrams, but partially captures 
the $multicenter$ character of each electron pair, especially in heavily stretched H$_2$. A remarkable improvement of BGE2 over PT2 and RPA in the H$_2$/H$_2^+$
challenge can be observed, which suggests that the one-electron ``self-correlation'' and the two-electron (near)-degeneracy static correlation can be
included simultaneously well at the second-order perturbation level in conjunction with a proper treatment of the multi-reference contributions of each 
electron pair. In addition of the size consistency, the advantage of the BGE2 correlation functional has been further demonstrated using H$_2$ in minimal basis.

However, for systems with large energy gaps or more electrons, BGE2 reduces to standard PT2 since the effect of the $e_{ab}$-coupling is 
nearly damped out. Further development on top of BGE2 could proceed as follows: 
1) From the semi-empirical double hybrid perspective, the BGE2 correlation formula could be a promising substitute of normal PT2, which 
opens an opportunity to extend the double-hybrid scheme into the realm of transition metals while keeping the accuracy 
achieved for main group elements\cite{igor:2009A}.
2) From the AC modeling perspective, we can improve the BGE2 model by satisfying more physical constraints. For example, in the strong 
correlation limit $\lambda\rightarrow\infty$, the BGE2 model depends asymptotically on $\lambda^{-1}$ rather than the correct 
$\lambda^{-1/2}$\cite{perdew:2000A,helgaker:2010A}. Promisingly, the plain sum-over-state PT2-like formula makes it easy to fulfill the correct
asymptotic behavior by introducing an additional term that depends on $\lambda^{-3/2}$ in the denominator of the BGE2 formula (Eq.\ \ref{Eq:BGE2}).
3) From the many-body perturbation theory perspective, it would be appealing to renormalize the BGE2 scheme in rPT2\cite{rinke:2013B},
which would be an alternative to construct advanced orbital-dependent functionals systematically.

Acknowledgments: IYZ thanks Professor Xin Xu for helpful discussion. Work at Aalto was supported by the Academy of Finland through its Centres of Excellence Programme (2012-2014 and 2015-2017) under project numbers 251748 and 284621.
\bibliography{paper}
\end{document}